\documentclass[physics,article,accept,oneauthor,pdftex]{Definitions/mdpi}

\firstpage{1} 
\pubvolume{1}
\issuenum{1}
\articlenumber{0}
\pubyear{2021}
\copyrightyear{2021}
\datereceived{} 
\dateaccepted{} 
\datepublished{} 
\hreflink{https://doi.org/} 

\usepackage{amssymb}

\newcommand{\pd}[1]{\, \partial #1 \,}

\newcommand{\g}{\ensuremath{\gamma}}
\newcommand{\zred}{z_{\rm red}}

\newcommand{\obs}{^{\rm obs}}

\newcommand{\E}[1]{\times 10^{#1}}
\Title{Studying the influence of external photon fields on blazar spectra using a one-zone hadro-leptonic time-dependent model}

\TitleCitation{Studying the influence of external photon fields on blazar spectra using a one-zone hadro-leptonic time-dependent model}

\Author{Michael Zacharias $^{1,2}$\orcidA{}}

\AuthorNames{Michael Zacharias}

\AuthorCitation{Zacharias, M.}

\address{%
$^{1}$ \quad Laboratoire Univers et Théories, Observatoire de Paris, Université PSL, CNRS, Université de Paris, 92190 Meudon, France; michael.zacharias@obspm.fr\\
$^{2}$ \quad Centre for Space Science, North-West University, Potchefstroom, 2520, South Africa; mzacharias.phys@gmail.com}

\corres{Correspondence: mzacharias.phys@gmail.com}

\abstract{The recent associations of neutrinos with blazars require the efficient interaction of relativistic protons with ambient soft photon fields. However, along side the neutrinos \g-ray photons are produced which interact with the same soft photon fields producing electron-positron pairs. The strength of this cascade has significant consequences on the photon spectrum in various energy bands and puts severe constraints on the pion and neutrino production. In this study, we discuss the influence of the external thermal photon fields (accretion disk, broad-line region, and dusty torus) on the proton-photon interactions employing a newly developed time-dependent one-zone hadro-leptonic code (\textit{OneHaLe}). We present steady-state cases, as well as a time-dependent case, where the emission region moves through the jet. Within the limits of this toy study, the external fields can disrupt the ``usual'' double-humped blazar spectrum. Similarly, a moving region would cross significant portions of the jet without reaching the previously-found steady states.}


\begin{document}

\section{Introduction}
The theory of blazar emission was transformed in the early 1990's by the introduction of the so-called \textit{external-Compton} scenario. The scenario explains the high-energy component of the spectral energy distribution (SED) through relativistic electrons inverse-Compton (IC) scattering soft, thermal photon fields that originate outside the jet. This transformation of blazar research was significantly driven by the works of Prof. Reinhard Schlickeiser and collaborators employing the accretion disk (AD) as a source for soft external photons \cite{dsm92,ds93,dss97,ds02}.

Blazars, a sub-class of active galaxies, are indeed peculiar objects with -- in the words of Prof. Schlickeiser \cite{ds92} -- 
\begin{quote}
properties [that] include high optical polarization, extreme optical variability, flat-spectrum radio emission associated with a compact core, and apparent superluminal motion. Such properties are thought to be produced by those few, rare extragalactic radio galaxies and quasars that are favorably aligned to permit us to look almost directly down a relativistically outflowing jet of matter expelled from a supermassive black hole.
\end{quote}
And despite the decades of research that have passed since the advent of the external-Compton model, a clear consensus on the source of the \g-ray emission in blazars is not yet reached. 

The low-energy component of the famous double-humped SED is the least controversial part, as synchrotron emission of relativistic electrons fits all the required properties (including the aforementioned polarization). However, the nature of the high-energy component is subject of intensive discussions. Within leptonic models, it is explained through IC emission -- either with the self-made synchrotron photons (synchrotron-self Compton, SSC) or with external photons, such as the AD, photons from the broad-line region (BLR) \cite{sbr94}, the dusty torus (DT) \cite{bsmm00,aps02}, or even the cosmic microwave background (CMB) \citep{bdf08,zw16}. However, if relativistic jets are also capable of accelerating protons, the \g-rays could also originate from such interactions, through direct proton-synchrotron emission, or via proton-photon interactions causing a cascade of pairs \cite{mb92,m93,muecke+00,dmpr12,petropouloumastichiadis15}. Especially, the production of pions would have the capability to discriminate between the leptonic and the hadronic scenario, as it would also produce neutrinos. While neutrinos have been associated with blazars recently \citep{icecube+18,hovatta+21}, the significances are not yet sufficient to claim a real detection. Nonetheless, the discussion is ongoing \cite{rbb19}, and the upcoming neutrino observatories KM3NET and IceCube-Gen2 may provide the definitive answer.

In this study, the hadro-leptonic model is combined with the external soft photons, to study their influence on the resulting pair cascade and the jet emission. A newly developed time-dependent, one-zone hadro-leptonic code -- \textit{OneHaLe} -- will be introduced (Section 2). It is used in Section 3 to study the influence of the external photon fields by first calculating steady-state spectra at various locations within the jet, as the region of influence of the soft photon fields on the jet is strongly distance-dependent. Subsequently, we present the case of an emission region moving outward passing through the various external photon fields. We note that the study conducted is a toy model: In order to properly identify the influence of the external fields, all other parameters of the emission region remain the same irrespective of the location. This may have significant consequences for the emerging spectra. Section 4 provides the discussion of the results and the conclusions. In the following, quantities in the host galaxy frame are marked with a hat, while quantities in the observer's frame are marked by the superscript ``obs''. Unmarked quanitites are either in the comoving frame of the emission region or invariant.
 
\section{Code description}
The code is based on the recently developed extended hadro-leptonic steady-state code \textit{ExHaLe-jet} \cite{zacharias+21}. In fact, the fundamental equations governing the particle and radiation processes are the same, and we only provide a brief overview here describing the free parameters.

We assume a spherical emission region with radius $R$ located a distance $z_0$ from the black hole within the jet, pervaded by a tangled magnetic field of strength $B$. The emission region moves with bulk Lorentz factor $\Gamma$ under a viewing angle $\theta^{\rm obs}$ with respect to the observer's line-of-sight implying a Doppler factor $\delta = [\Gamma(1-\beta_{\Gamma}\cos{\theta^{\rm obs}})]^{-1}$. Here, $\beta_{\Gamma}=\sqrt{1-\Gamma^{-2}}$.

The Fokker-Planck equation governing the time-dependent evolution of a given particle species $i$ (protons, charged pions, muons, or electrons) with spectral density $n_i(\chi)$ is given as

\begin{align}
     \frac{\pd{n_i(\chi,t)}}{\pd{t}} &= \frac{\pd{}}{\pd{\chi}} \left[ \frac{\chi^2}{(a+2)t_{\rm acc}} \frac{\pd{n_i(\chi,t)}}{\pd{\chi}} \right] \nonumber \\
	 &\quad - \frac{\pd{}}{\pd{\chi}} \left( \dot{\chi}_i n_i(\chi,t) \right) + Q_i(\chi,t)
	 - \frac{n_i(\chi,t)}{t_{\rm esc}} - \frac{n_i(\chi, t)}{\gamma t^{\ast}_{i,{\rm decay}}}. 
	 \label{eq:fpgen}
\end{align}
For numerical reasons, we use the normalized particle momentum $\chi=p_i/(m_i c)=\gamma\beta$, where $m_i$ is the particle mass, $c$ the speed of light, $\gamma$ the particle's Lorentz factor, and $\beta=\sqrt{1-\gamma^{-2}}$. The first term on the right-hand side of Eq.~(\ref{eq:fpgen}) describes Fermi-II acceleration through scattering of particles on magnetohydrodynamic waves. We use the parametrization of \cite{weidingerspanier15} with $a=9v_s^2/4v_{A}^2$, $v_{s/A}$ the shock and Alfv\`{e}n speed, respectively, and the energy-independent acceleration time scale $t_{\rm acc}$. This parametrization approximates the momentum diffusion through hard-sphere scattering.

The second term on the right-hand side of Eq.~(\ref{eq:fpgen}) provides momentum changes $\dot{\chi}_i$ through gains (Fermi-I acceleration $\dot{\chi}_{\rm FI} = \chi/t_{\rm acc}$) and continuous losses. All charged particles lose energy through synchrotron radiation and adiabatic expansion of the emission region. Protons also lose energy through pion production \footnote{The pion production can turn a proton into a neutron. As we do not explicitly consider neutrons at this point, we approximate this effect by a continuous loss process instead of a catastrophic loss. This channel is marked as ``neutron'' losses in Figs.~\ref{fig:par_ste} and \ref{fig:par_mot}, while the nominal pion production cooling term is marked as ``pion''.} and Bethe-Heitler pair production, while electrons suffer additional losses through IC scattering of ambient photon fields. These ambient fields consist of all intrinsically produced radiation fields -- such as synchrotron -- as well as the external photon fields, namely the AD, the BLR, the DT, and the CMB. 

The remaining three terms on the right-hand side of Eq.~(\ref{eq:fpgen}) mark the injection of particles, the escape of particles from the emission region, and the decay of unstable particles, respectively. $t^{\ast}_{i,{\rm decay}}$ is the proper decay time scale, which is $2.6\times 10^{-8}\,$s for charged pions, and $2.2\times 10^{-6}\,$s for muons, respectively. As neutral pions decay after $2.8\times 10^{-17}\,$s into \g\ rays, we do not solve Eq.~(\ref{eq:fpgen}) for neutral pions, but calculate their radiation output directly from their injection spectrum.

While we consider Fermi-I and II acceleration terms, we treat them merely as re-acceleration processes characterized by the acceleration time scale $t_{\rm acc}=\eta_{\rm acc}t_{\rm esc}$; namely, a multiple $\eta_{\rm acc}$ of the escape time scale \cite{diltz+15}. We do not consider the primary acceleration of protons and electrons, which may take place in small sub-regions of the larger emission region \citep{weidingerspanier15,chen+15}, but approximate it through the injection term $Q(\chi,t)$. Here, we use a simple power-law injection with spectral index $s_i$ between a minimum and maximum Lorentz factor, $\gamma_{\rm min,i}$ and $\gamma_{\rm max,i}$, respectively. The injection normalization $Q_{0,i}(t)$ is given by

\begin{align}
	Q_{0,i}(t) = \frac{L_{\rm inj,i}(t)}{Vm_ic^2}\begin{cases}
	                                          	\frac{2-s_i(t)}{\gamma_{\rm max,i}^{2-s_i(t)}-\gamma_{\rm min,i}^{2-s_i(t)}} & \mbox{if}\ s_i(t) \neq 2 \\
	                                          	\left( \ln{\frac{\gamma_{\rm max,i}}{\gamma_{\rm min,i}}} \right)^{-1} & \mbox{if}\ s_i(t) = 2
	                                          \end{cases}
\label{eq:inj},
\end{align}
with the injection luminosity $L_{\rm inj,i}$ and the volume $V$ of the spherical emission region. The injection functions for pions and muons are calculated directly from the photo-hadron interactions \cite{huemmer+10} and decays. We emphasize again that Eq.~(\ref{eq:fpgen}) is explicitly solved for (charged) pions and muons.

The escape of particles is described by $t_{\rm esc} = \eta_{\rm esc}R/c$, a multiple $\eta_{\rm esc}$ of the light travel time. As $\eta_{\rm esc}>1$, this mimics the advective flow of particles through the emission region.

Equation (\ref{eq:fpgen}) is solved with a Chang\&Cooper routine \cite{changcooper70}. For a detailed description, see \cite{dmytriiev+21,zacharias+21}.

The interaction of protons with photons can result in the creation of pions. Charged pions decay into muons, which in turn decay into electrons. During both decay processes, neutrinos are produced. The neutrino spectra are calculated following \cite{barr+88,gaisser90,zacharias+21}. The secondary electrons produced in this decay chain are injected into the electron-Fokker-Planck equation along with the primary electrons. Additionally, secondary electrons are also produced from Bethe-Heitler pair production and \g-\g\ pair production.

We do not consider explicitly neutrons in this code. Their number density is low compared to the proton density \cite{muecke+00}; so their effect is small. Nonetheless, we plan to rectify this issue in a future update of the code.

The photon density $n_{\rm ph}$ within the emission region is governed by the radiation transport equation:

\begin{align}
	\frac{\pd{n_{\rm ph}(\nu,t)}}{\pd{t}} = \frac{4\pi}{h\nu} j_{\nu}(t) - n_{\rm ph}(\nu,t) \left( \frac{1}{t_{\rm esc,ph}} + \frac{1}{t_{\rm abs}} \right) 
	\label{eq:radtrans}.
\end{align}
with the emissivity $j_{\nu}$, the photon escape time scale $t_{\rm esc,ph}=4R/3c$, and the absorption time scale $t_{\rm abs}$ due to synchrotron-self absorption and $\gamma$-$\gamma$ pair production. For the latter, all internal and external photon fields are considered. From the photon distribution $n_{\rm ph}$, we can calculate the spectral luminosity in the observer's frame

\begin{align}
	\nu\obs L\obs_{\nu\obs} = \delta^4 \frac{h\nu^2V}{t_{\rm esc,ph}} n_{\rm ph}(\nu,t)
	\label{eq:speclum}.
\end{align}
Equations (\ref{eq:radtrans}) and (\ref{eq:speclum}) hold for all radiation processes within the emission region.

On their way from the source to the observer, \g-ray photons are subject to further absorption processes. We consider the important cases of absorption in the BLR and DT following the prescription in \cite{be16}.

The AD is described with a standard Shakura-Sunyaev disk \cite{shakurasunnyayev73} implying that the disk is fully described through the mass of the supermassive black hole $M_{\rm BH}$ and its accretion efficiency $\eta_{\rm SS}$ (or Eddington ratio). The proper transformation of the angles into the comoving frame is considered. The BLR and DT are approximated as isotropic photon fields in the host galaxy frame within an distance $\hat{R}_{\rm BLR}$ and $\hat{R}_{\rm DT}$ from the black hole, and their energy distribution is given through a grey-body spectrum of temperature $\hat{T}_{\rm BLR}$ and $\hat{T}_{\rm DT}$ normalized to a luminosity of $\hat{L}_{\rm BLR}$ and $\hat{L}_{\rm DT}$, respectively.

The above description holds for both steady-state and time-dependent cases. The steady state is achieved if the proton and electron densities derived from Eq.~(\ref{eq:fpgen}), do not vary by more than $10^{-4}$ compared to the respective values of the previous two time steps. Time-dependency can be achieved by varying any of the free parameters, in which case steady states may not be achieved from time step to time step.

\section{Influence of the external fields}
\begin{specialtable}[t] 
\small
\caption{Free parameters of the code along with symbols, units and toy model values. 
\label{tab:freeparam}}
\begin{tabular}{lccc}
\toprule
\textbf{Parameter}	& & \textbf{Unit}	& \textbf{Value}\\
\midrule
Redshift						& $\zred$				& 				& $0.536$\\
Location in jet					& $z_0$					& cm			& $1.0\E{16}$, $1.0\E{17}$, $1.0\E{18}$, $1.0\E{19}$\\
Magnetic field					& $B$					& G				& $50$\\
Radius							& $R$					& cm			& $4.5\E{15}$\\
Bulk Lorentz factor				& $\Gamma$				& 				& $50$\\
Observation angle				& $\theta^{\rm obs}$	& deg			& $1.3$\\
Proton injection luminosity		& $L_{\rm inj,p}$		& erg/s			& $3.0\E{43}$\\
Proton spectral index			& $s_p$					& 				& $2.1$\\
Proton min Lorentz factor		& $\gamma_{\rm min,p}$	& 				& $4.0\E{5}$\\
Proton max Lorentz factor		& $\gamma_{\rm max,p}$	& 				& $2.5\E{8}$\\
Electron injection luminosity	& $L_{\rm inj,e}$		& erg/s			& $2.0\E{41}$\\
Electron spectral index			& $s_e$					& 				& $3.0$\\
Electron min Lorentz factor		& $\gamma_{\rm min,e}$	& 				& $5.0\E{1}$\\
Electron max Lorentz factor		& $\gamma_{\rm max,e}$	& 				& $2.0\E{3}$\\
Multiple escape time			& $\eta_{\rm esc}$		& 				& $5$\\
Multiple acceleration time		& $\eta_{\rm acc}$		& 				& $30$\\
Black hole mass					& $M_{\rm BH}$			& $M_{\odot}$	& $3.0\E{8}$\\
AD efficiency					& $\eta_{\rm SS}$		& 				& $0.08$\\
BLR luminosity					& $\hat{L}_{\rm BLR}$	& erg/s			& $2.3\E{44}$\\
BLR temperature					& $\hat{T}_{\rm BLR}$	& K				& $1.0\E{4}$\\
BLR radius						& $\hat{R}_{\rm BLR}$	& cm			& $7.6\E{16}$\\
DT luminosity					& $\hat{L}_{\rm DT}$	& erg/s			& $3.0\E{44}$\\
DT temperature					& $\hat{T}_{\rm DT}$	& K				& $5.0\E{2}$\\
DT radius						& $\hat{R}_{\rm DT}$	& cm			& $4.2\E{18}$\\
\bottomrule
\end{tabular}
\end{specialtable}
\begin{figure}[tbh]
\begin{minipage}{0.49\linewidth}
\centering \resizebox{\hsize}{!}
{\includegraphics{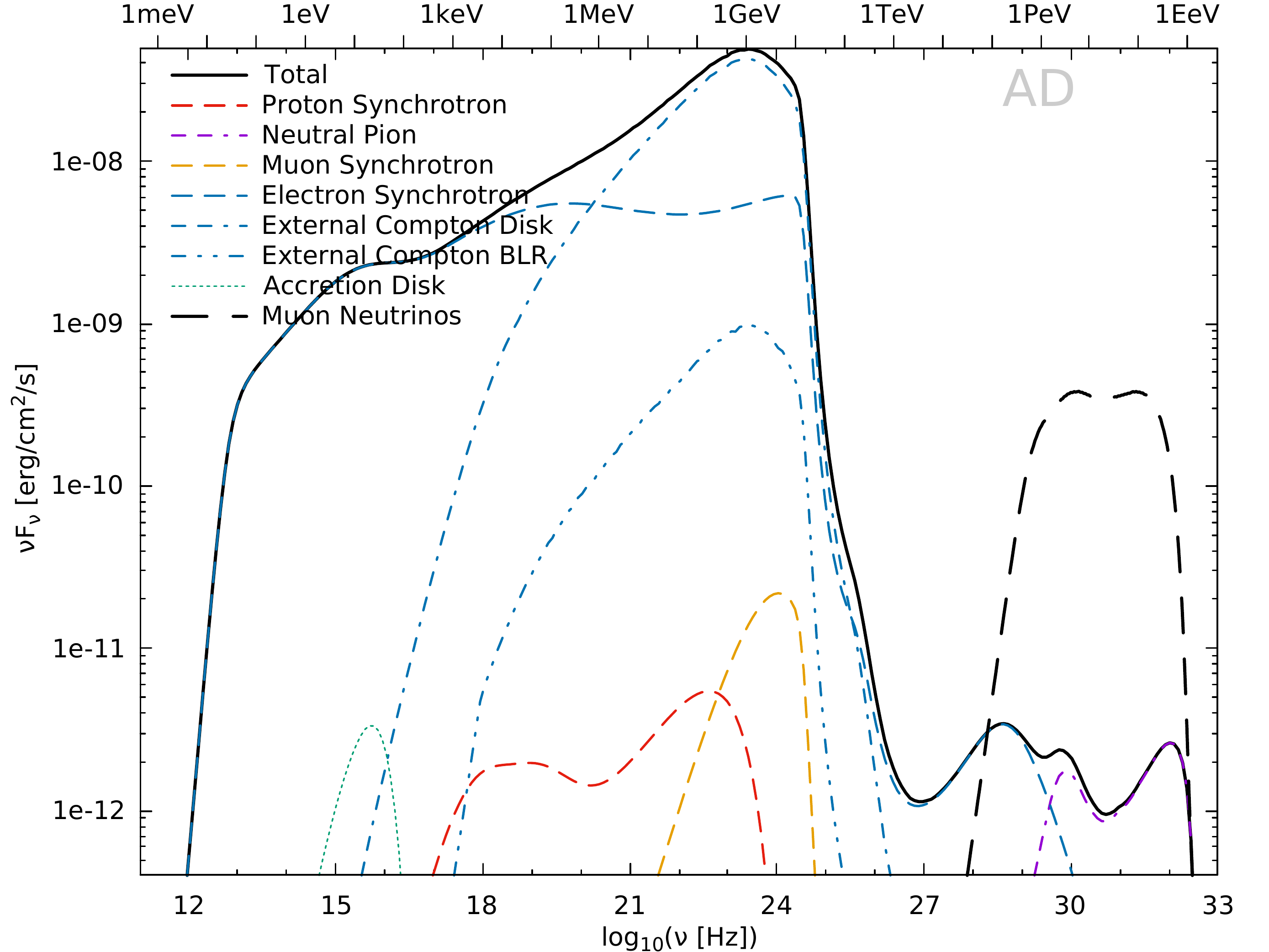}}
\end{minipage}
\hspace{\fill}
\begin{minipage}{0.49\linewidth}
\centering \resizebox{\hsize}{!}
{\includegraphics{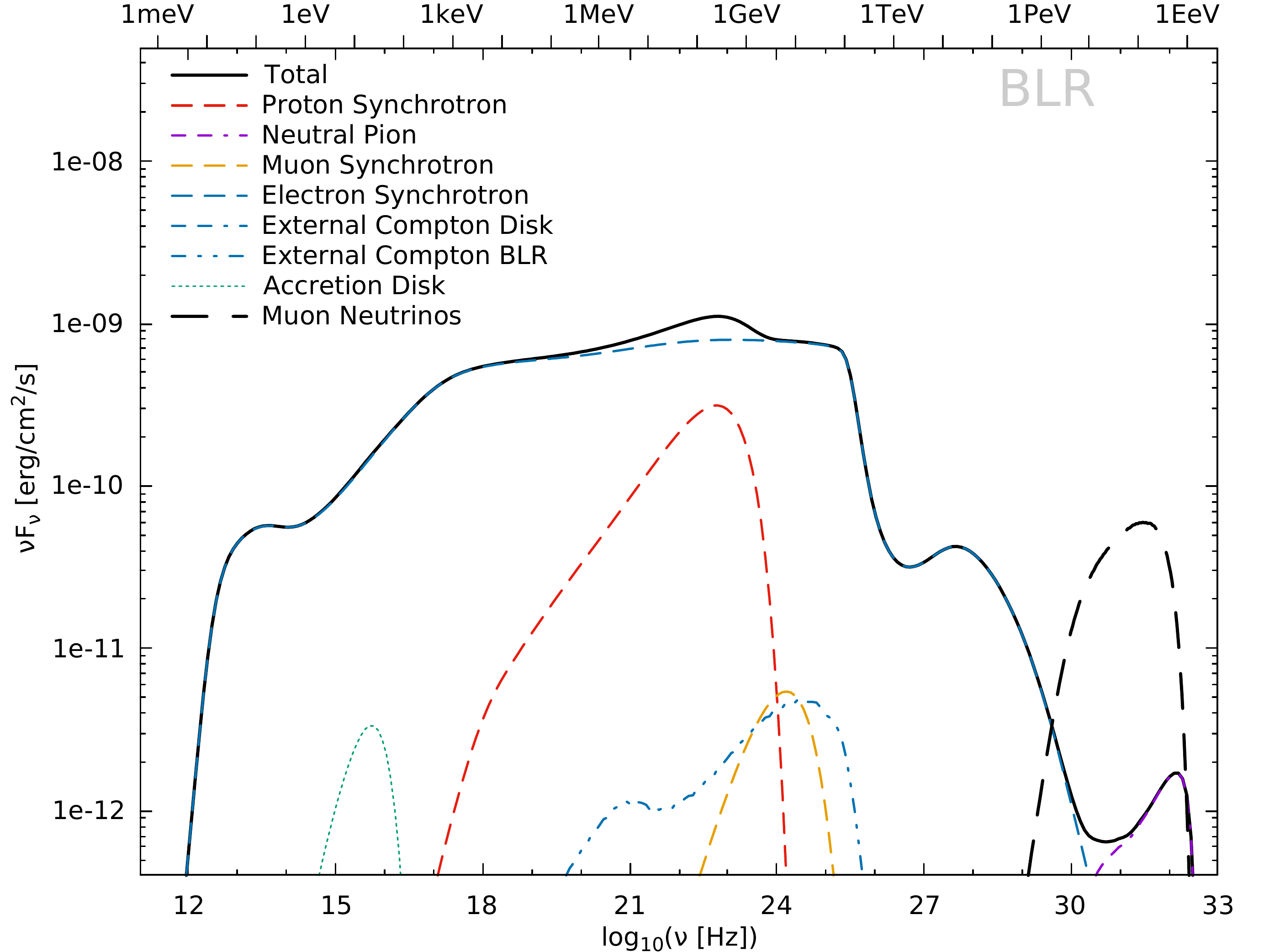}}
\end{minipage}
\newline
\begin{minipage}{0.49\linewidth}
\centering \resizebox{\hsize}{!}
{\includegraphics{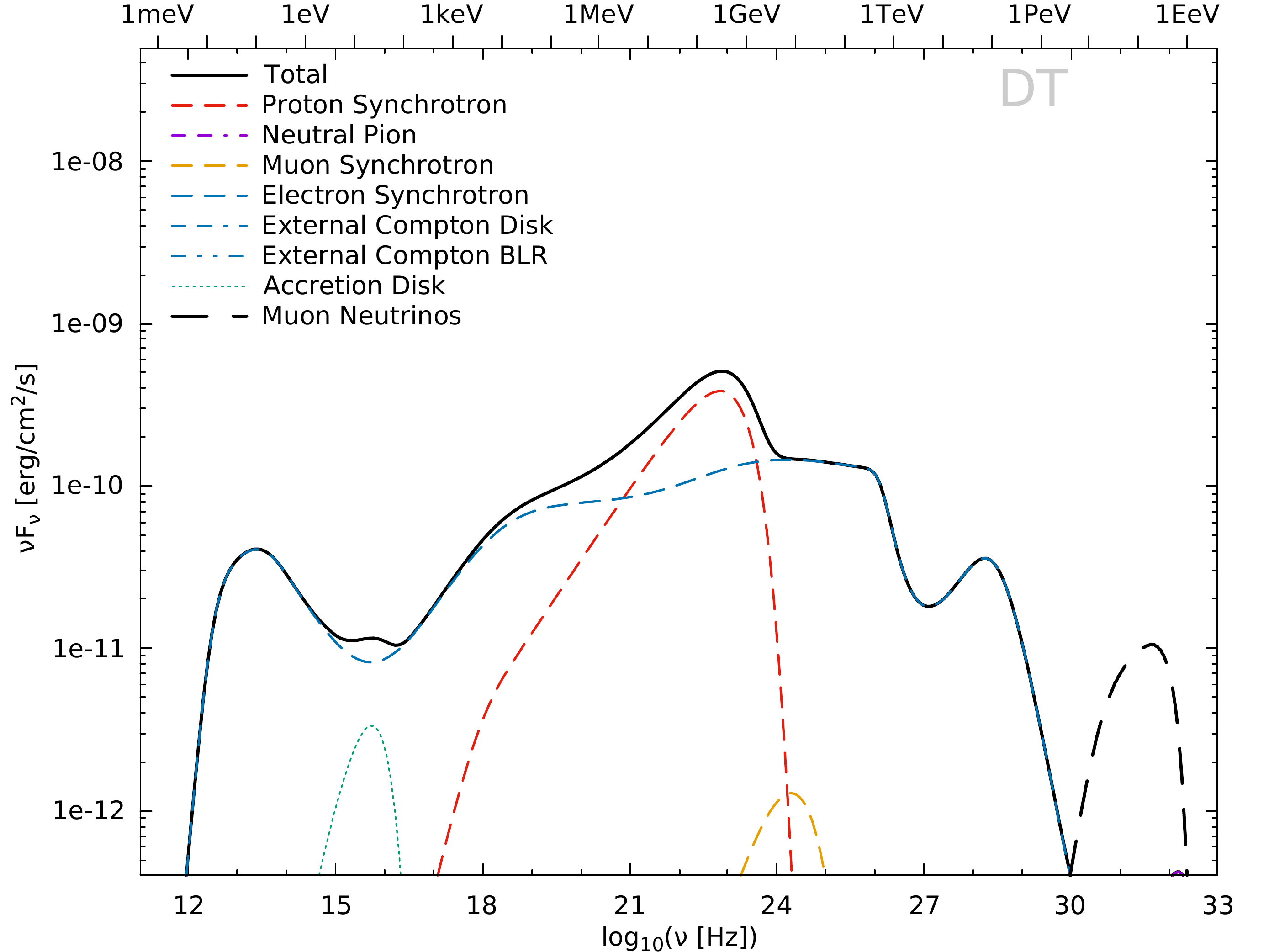}}
\end{minipage}
\hspace{\fill}
\begin{minipage}{0.49\linewidth}
\centering \resizebox{\hsize}{!}
{\includegraphics{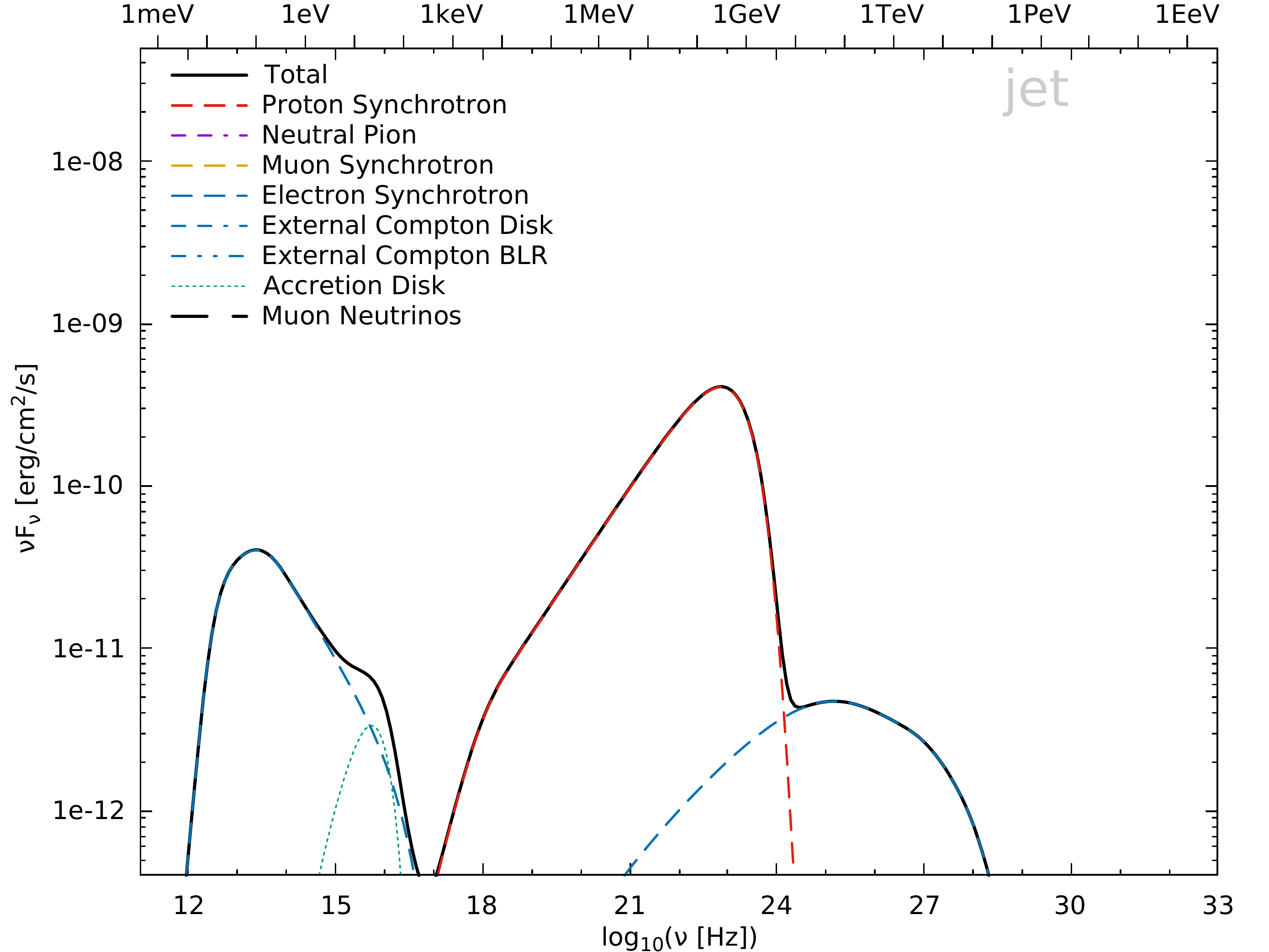}}
\end{minipage}
\caption{SEDs in the observer's frame for the four locations: close to the AD (top left), within the BLR (top right), within the DT (bottom left), and outside the external fields (jet, bottom right). The black solid line marks the total photon spectrum, while the colored lines mark individual photon components as labeled. Only those processes are labeled, which are visible in at least one panel. No external absorption is applied implying that photon spectra are shown as they leave the jet. The black dashed line marks the neutrino spectrum.
\label{fig:spe_ste}}
\end{figure}   
\begin{figure}[tbh]
\begin{minipage}{0.49\linewidth}
\centering \resizebox{\hsize}{!}
{\includegraphics{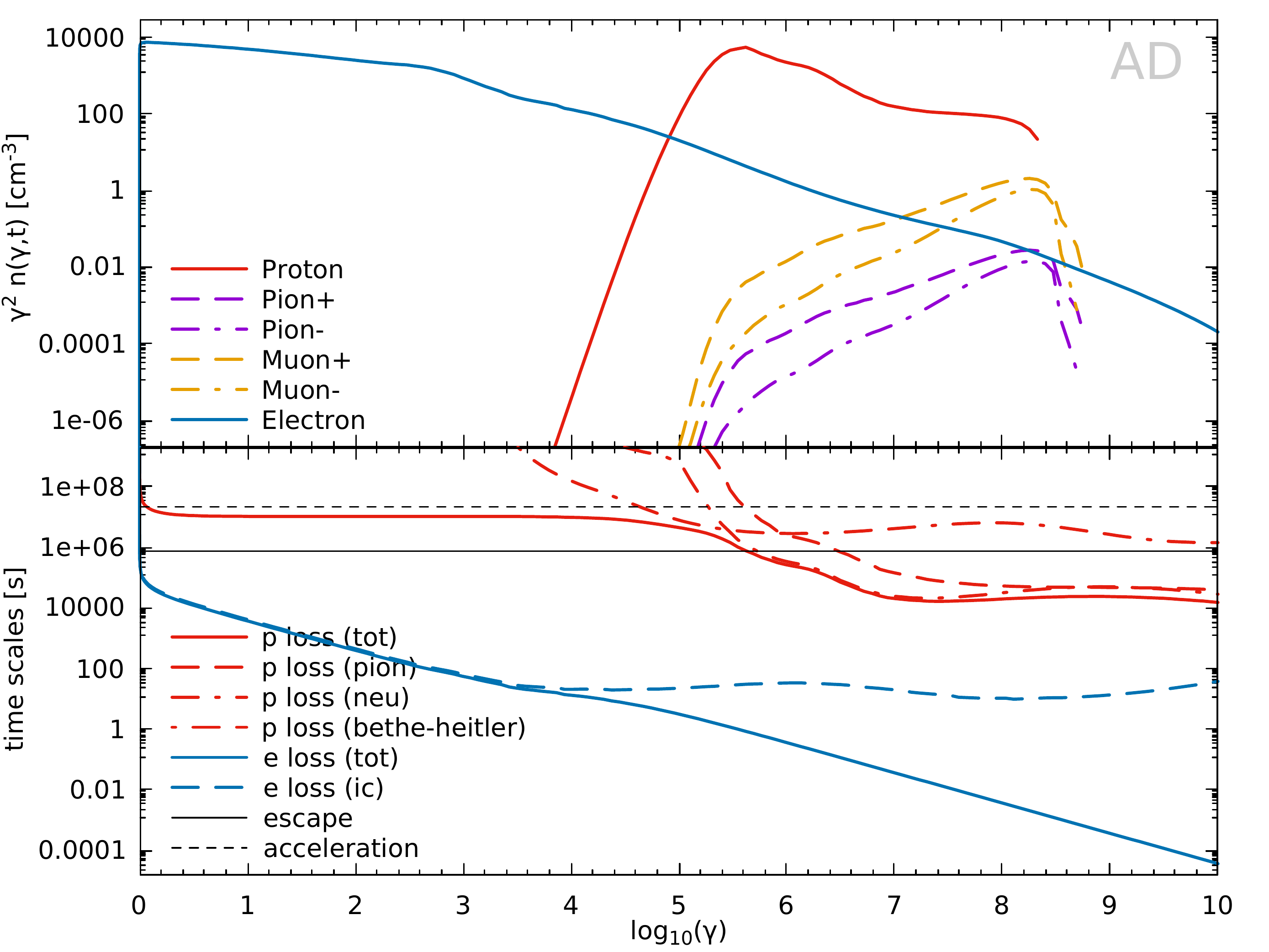}}
\end{minipage}
\hspace{\fill}
\begin{minipage}{0.49\linewidth}
\centering \resizebox{\hsize}{!}
{\includegraphics{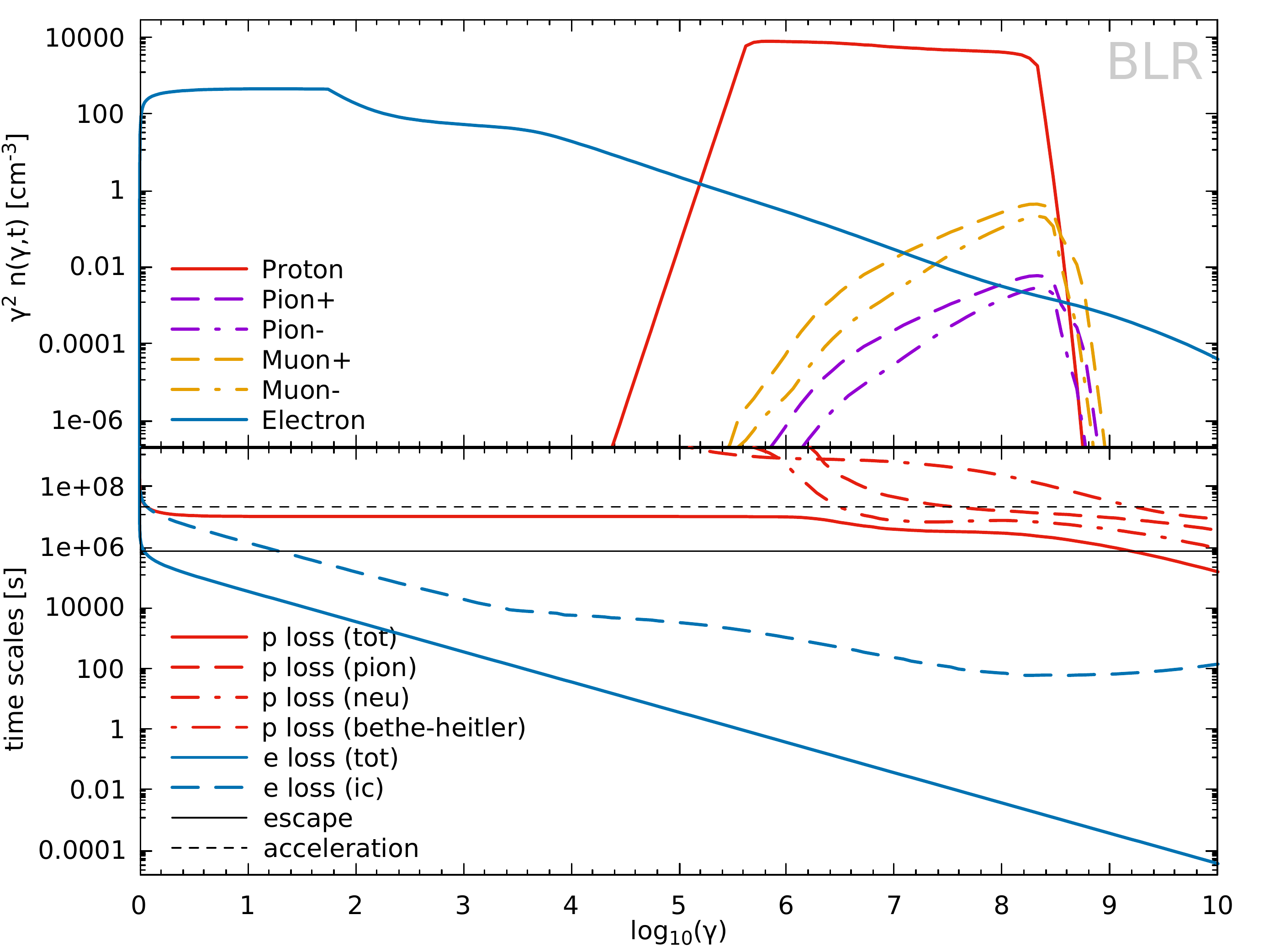}}
\end{minipage}
\newline
\begin{minipage}{0.49\linewidth}
\centering \resizebox{\hsize}{!}
{\includegraphics{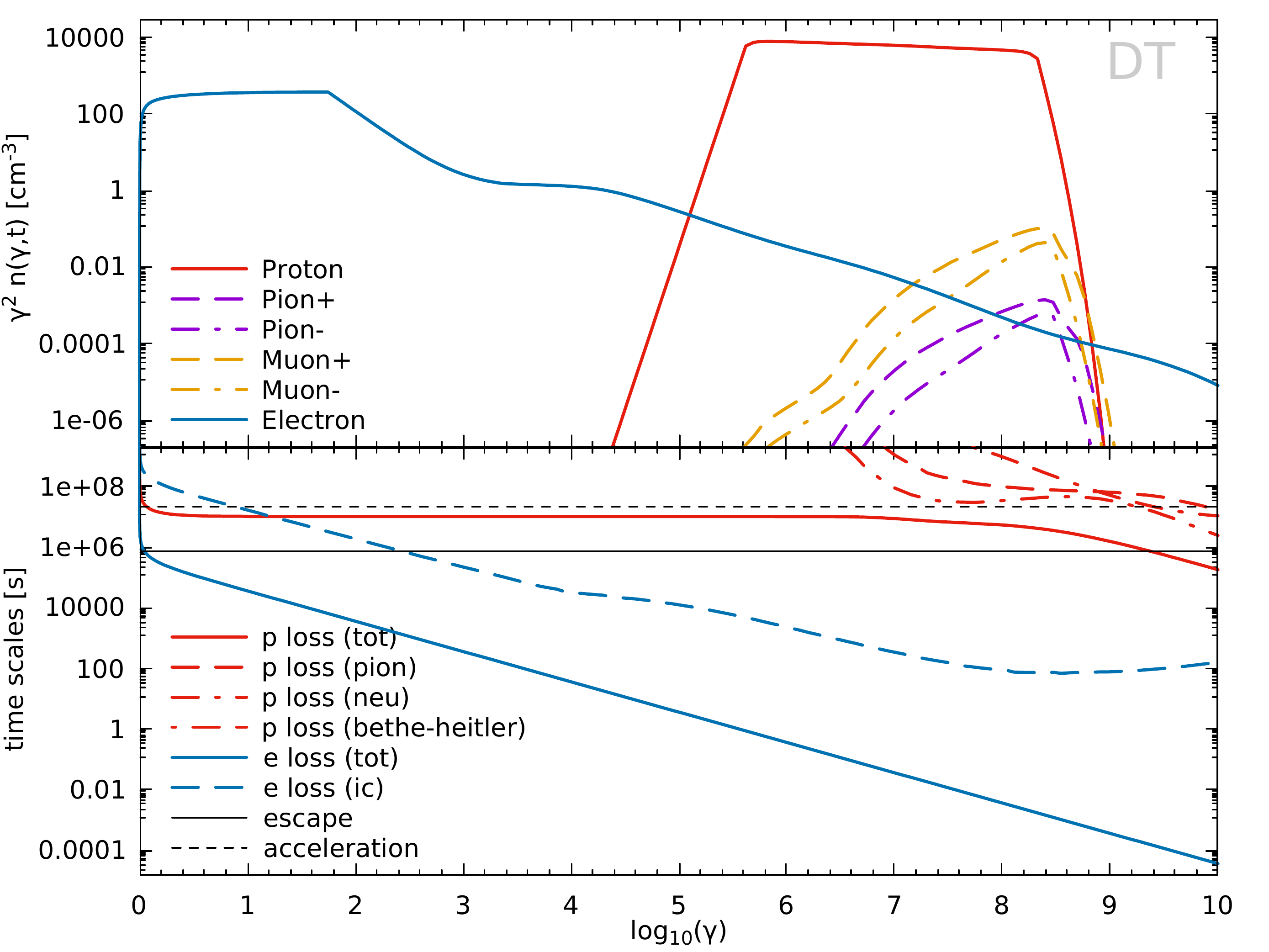}}
\end{minipage}
\hspace{\fill}
\begin{minipage}{0.49\linewidth}
\centering \resizebox{\hsize}{!}
{\includegraphics{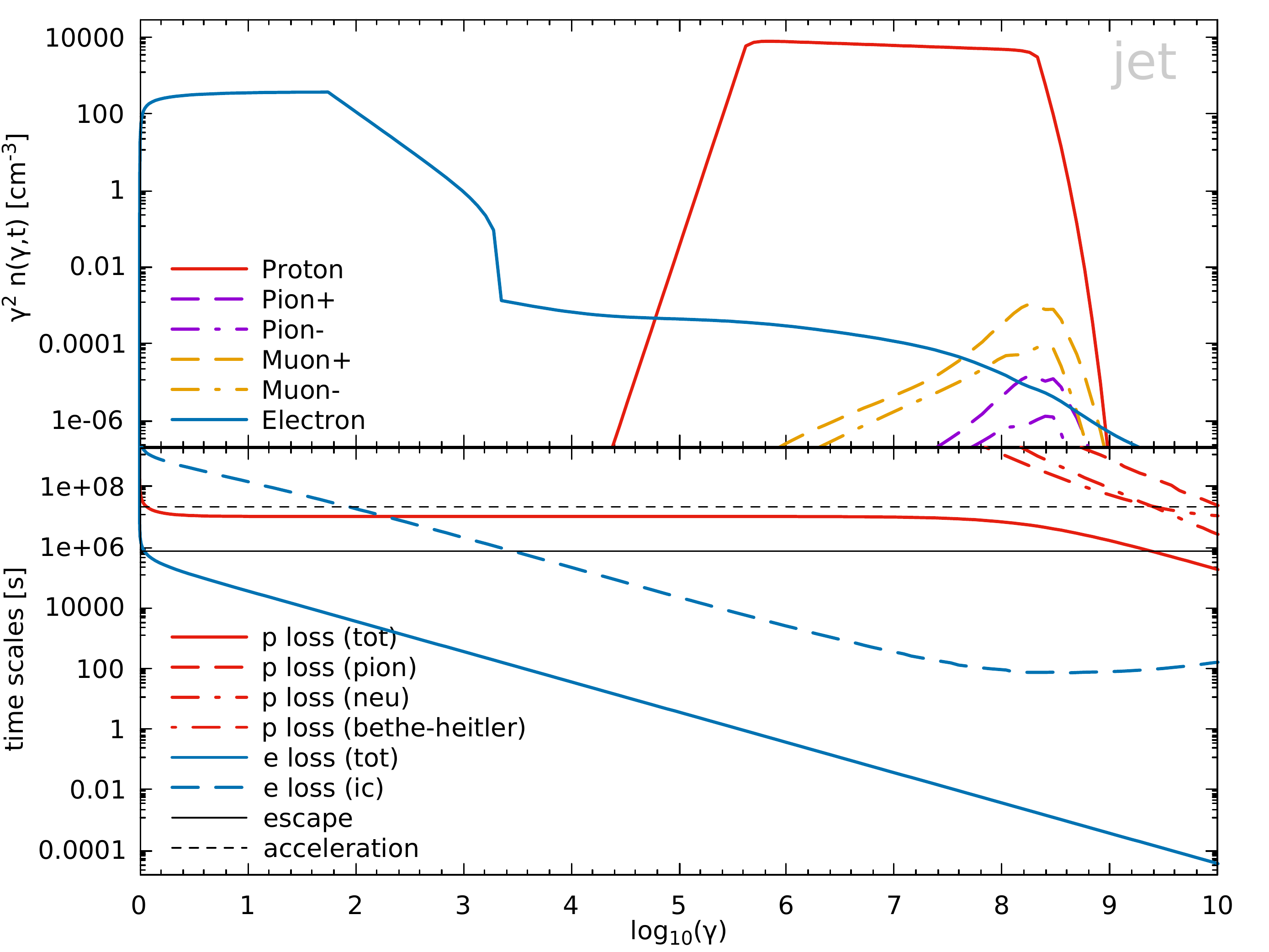}}
\end{minipage}
\caption{Steady-state particle distributions (times Lorentz factor squared), and relevant time scales as labeled as a function of Lorentz factor $\g$ for the same locations as in Fig.~\ref{fig:spe_ste}. For proton losses, the total, photo-pion, ``neutron'', and Bethe-Heitler loss time scales are shown. Adiabatic losses dominate at lower proton energies, where the loss time scale is constant, while synchrotron losses may contribute at the highest proton energies. For electron losses, the total, and IC loss time scales are shown. Synchrotron losses dominate, where IC losses are negligible, while adiabatic losses are irrelevant. 
\label{fig:par_ste}}
\end{figure}   
\begin{figure}[tbh]
\begin{minipage}{0.49\linewidth}
\centering \resizebox{\hsize}{!}
{\includegraphics{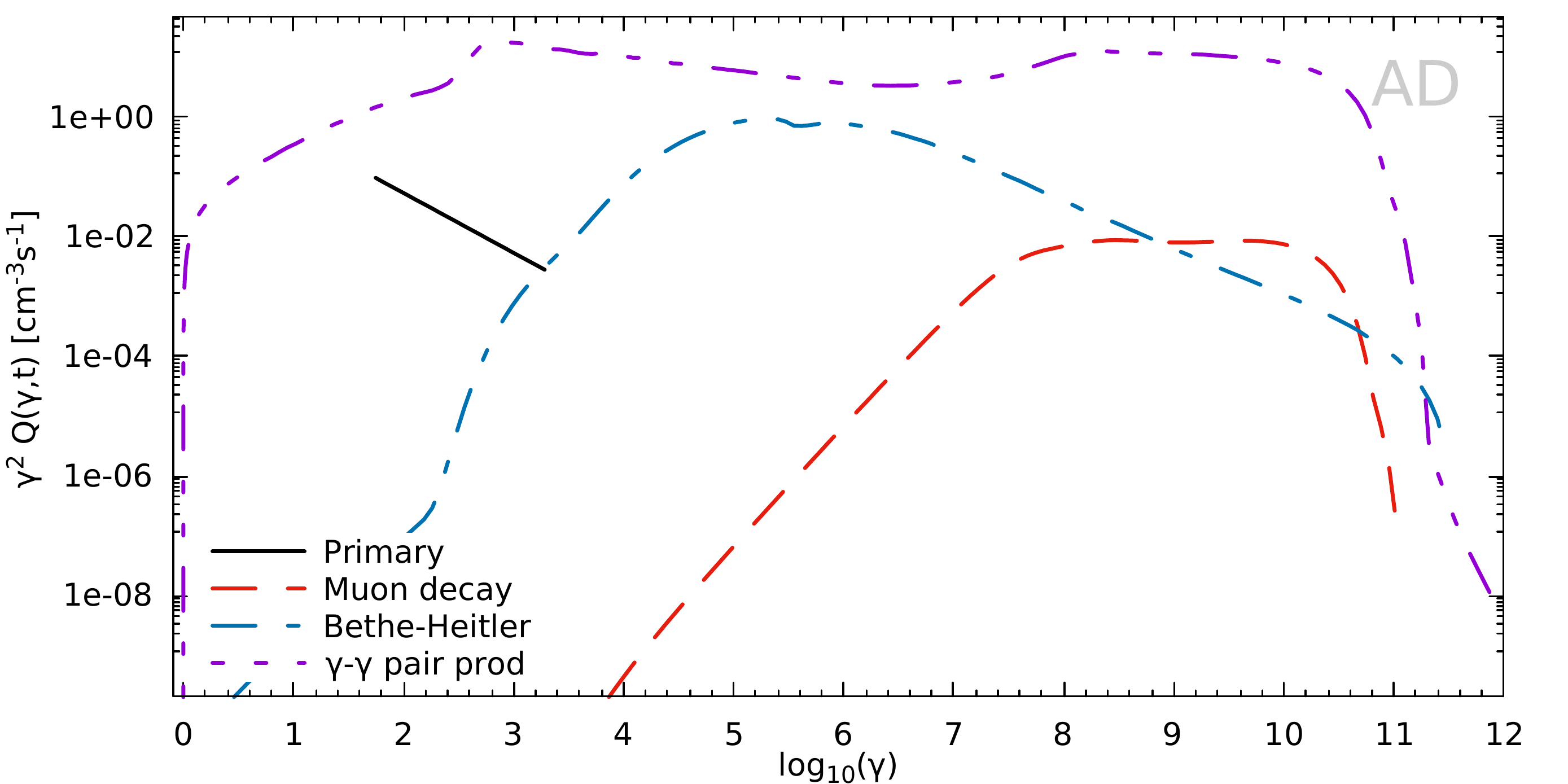}}
\end{minipage}
\hspace{\fill}
\begin{minipage}{0.49\linewidth}
\centering \resizebox{\hsize}{!}
{\includegraphics{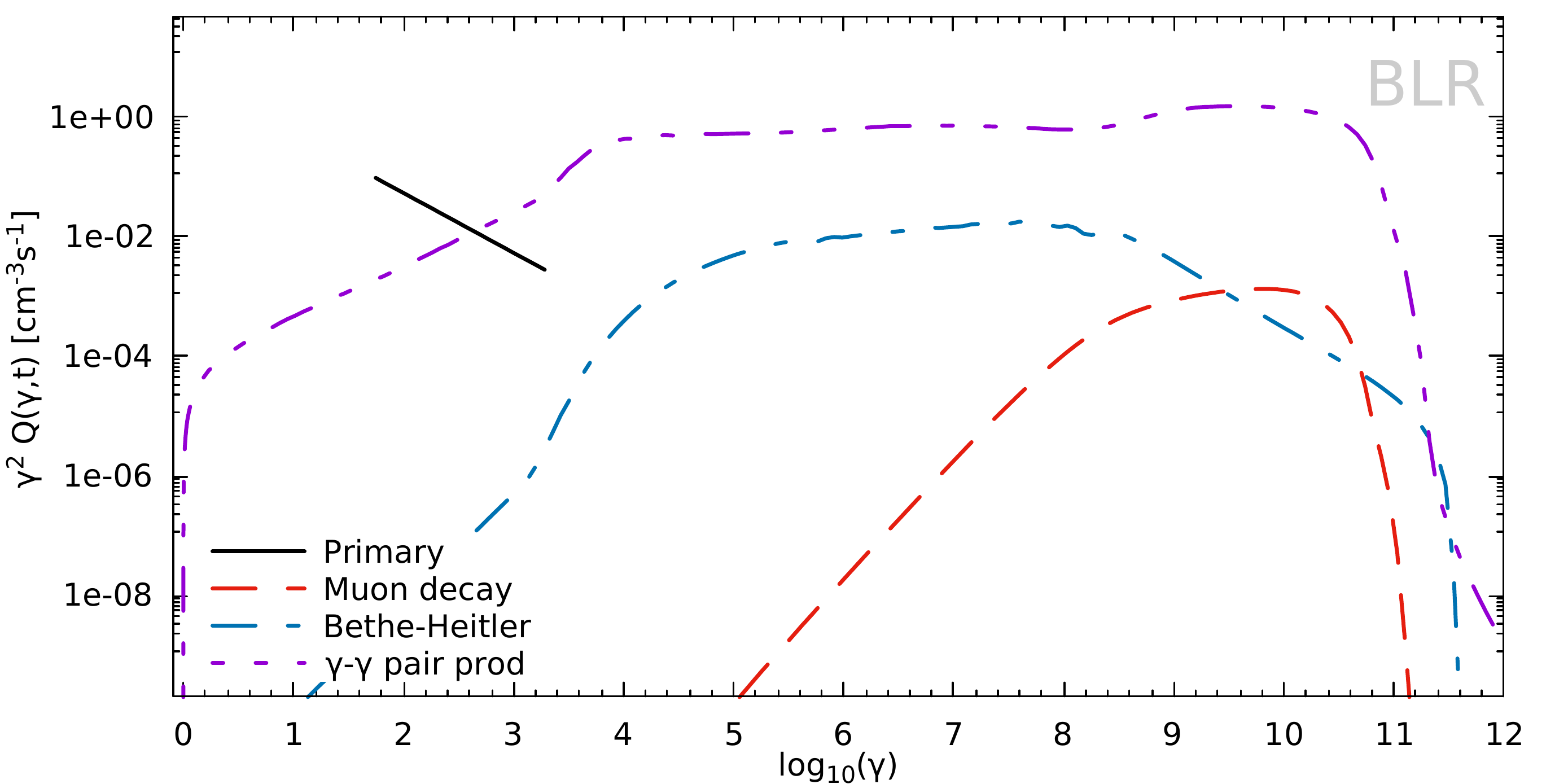}}
\end{minipage}
\newline
\begin{minipage}{0.49\linewidth}
\centering \resizebox{\hsize}{!}
{\includegraphics{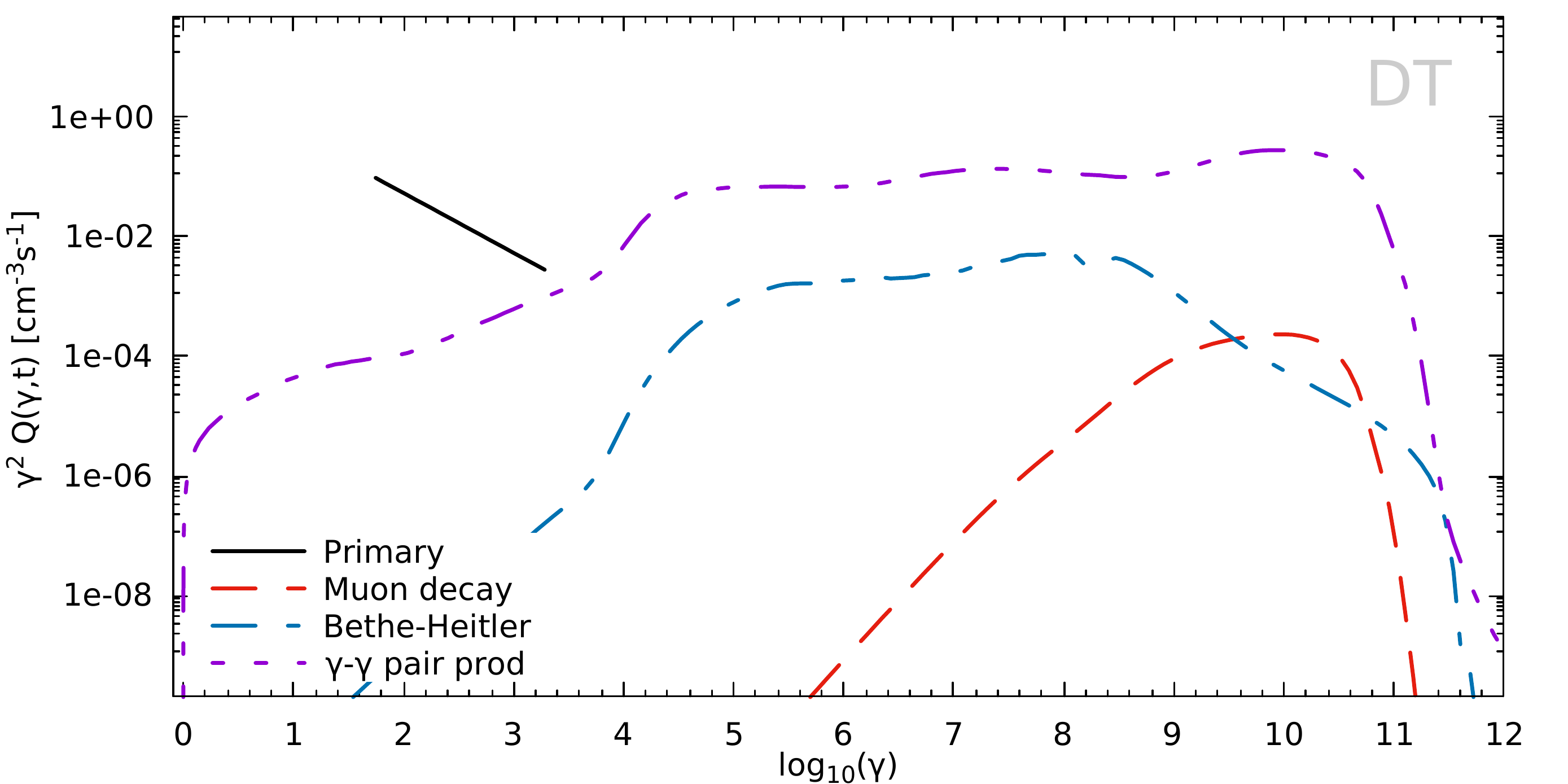}}
\end{minipage}
\hspace{\fill}
\begin{minipage}{0.49\linewidth}
\centering \resizebox{\hsize}{!}
{\includegraphics{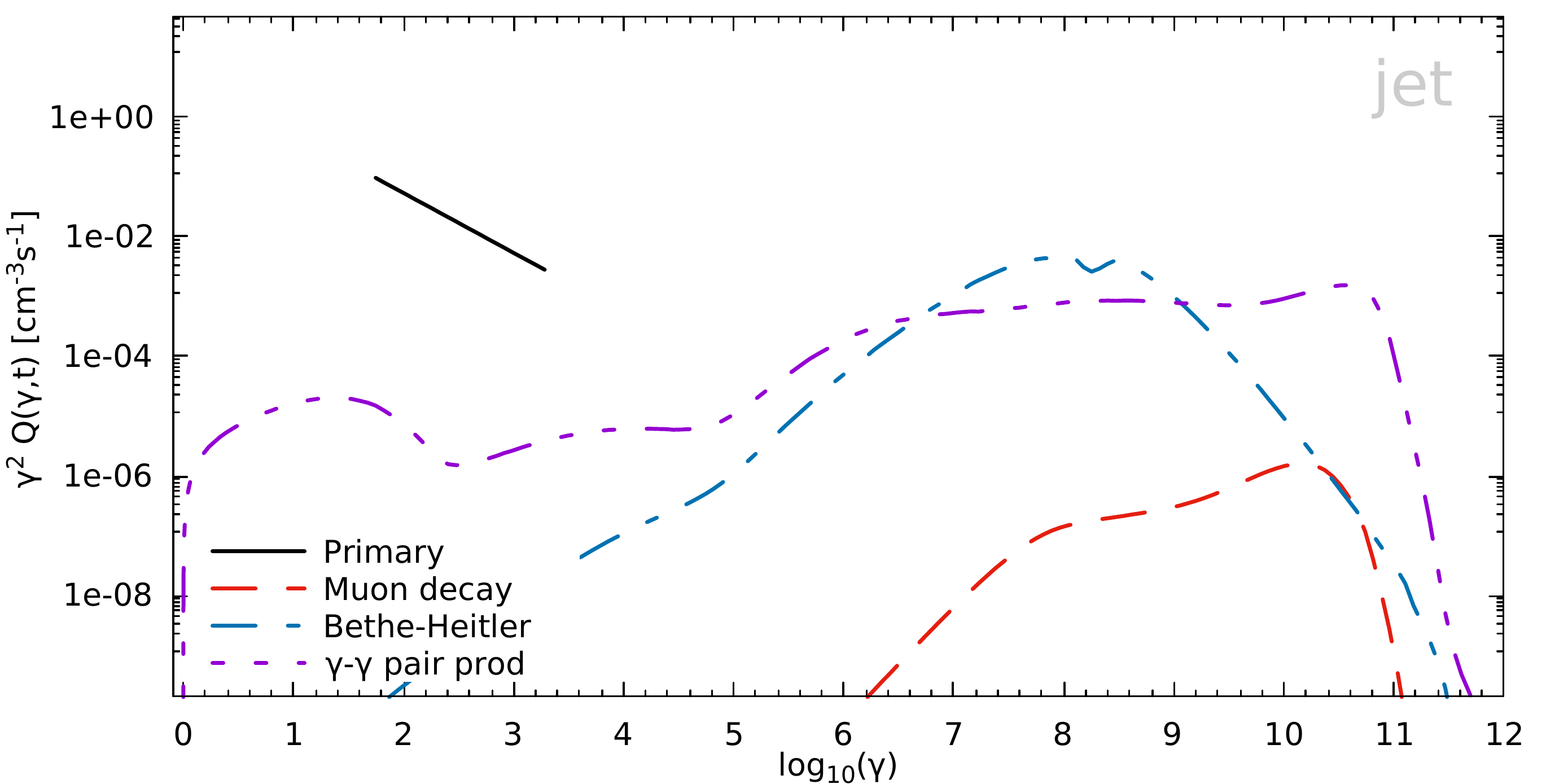}}
\end{minipage}
\caption{Steady-state electron injection rates $Q$ (times Lorentz factor squared) as a function of Lorentz factor $\g$ as labeled for the same locations as in Fig.~\ref{fig:spe_ste}. 
\label{fig:inj_ste}}
\end{figure}   
\begin{figure}[tbh]
\begin{minipage}{0.49\linewidth}
\centering \resizebox{\hsize}{!}
{\includegraphics{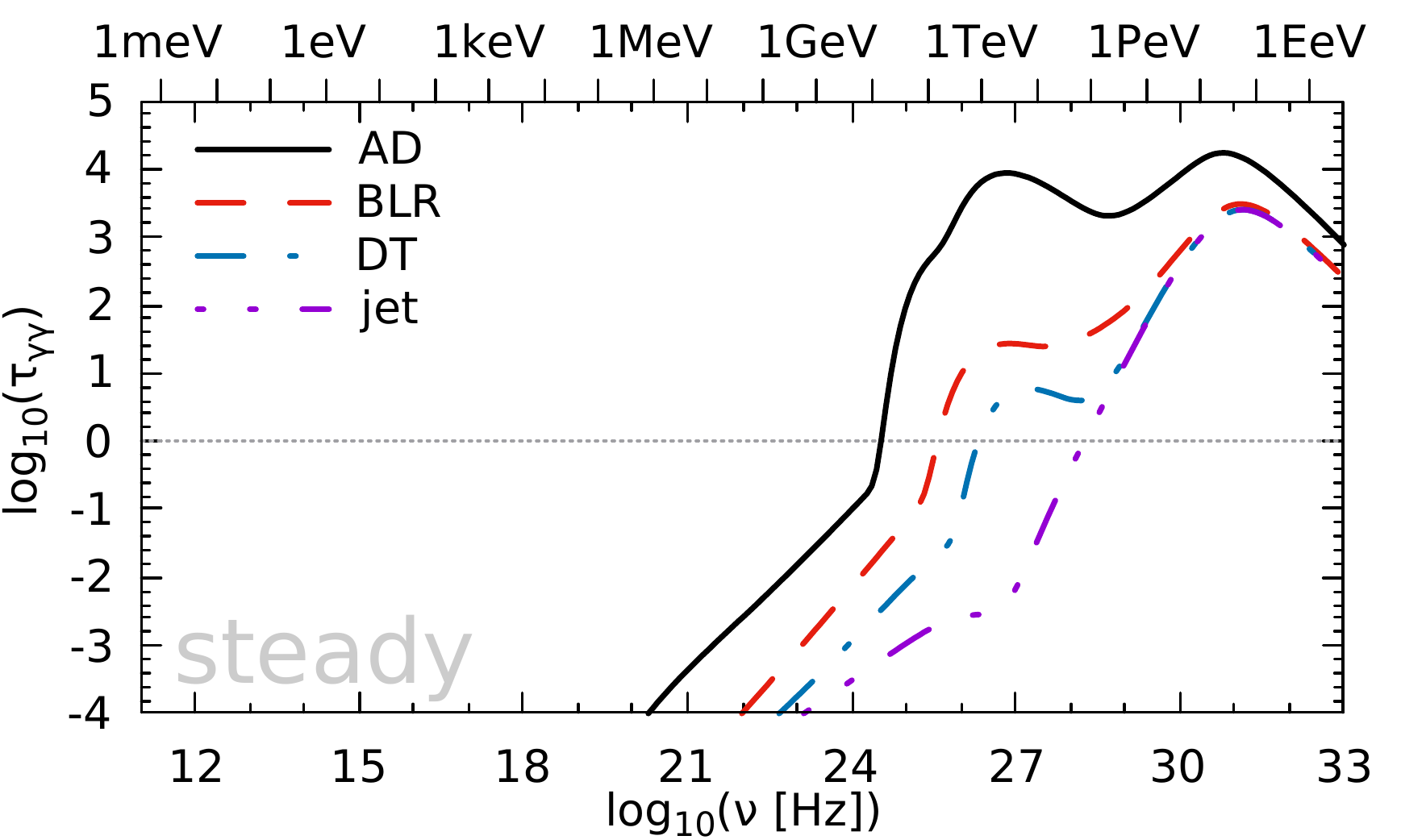}}
\end{minipage}
\hspace{\fill}
\begin{minipage}{0.49\linewidth}
\centering \resizebox{\hsize}{!}
{\includegraphics{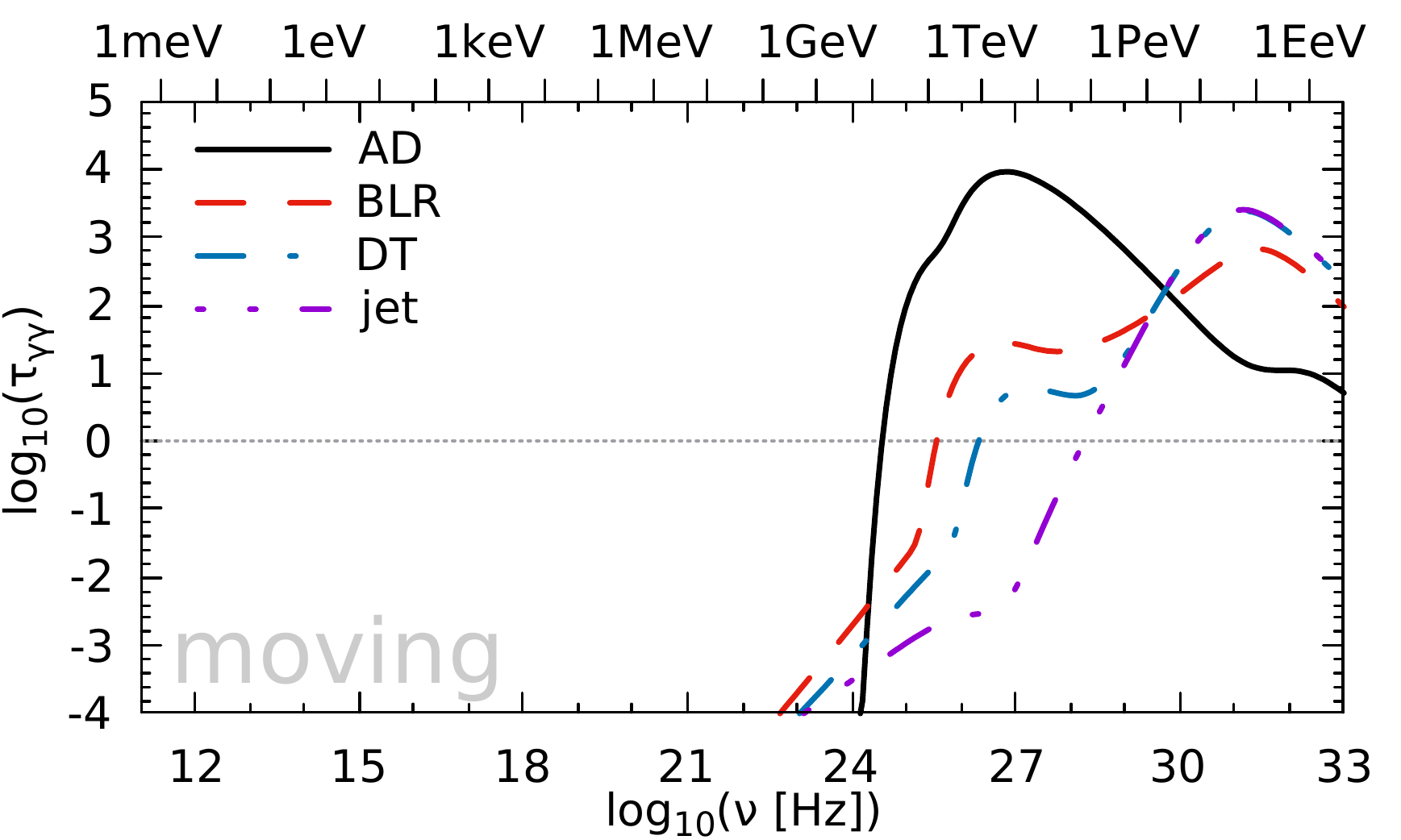}}
\end{minipage}
\caption{Optical depth $\tau_{\gamma\gamma}$ due to \g-\g\ pair production as a function of frequency (observer's frame) for the steady-state cases (left) and the moving-blob case (right) at the different positions within the jet as labeled. The thin horizontal line marks $\tau_{\gamma\gamma}=1$.  
\label{fig:abs}}
\end{figure}   
\begin{figure}[tbh]
\begin{minipage}{0.49\linewidth}
\centering \resizebox{\hsize}{!}
{\includegraphics{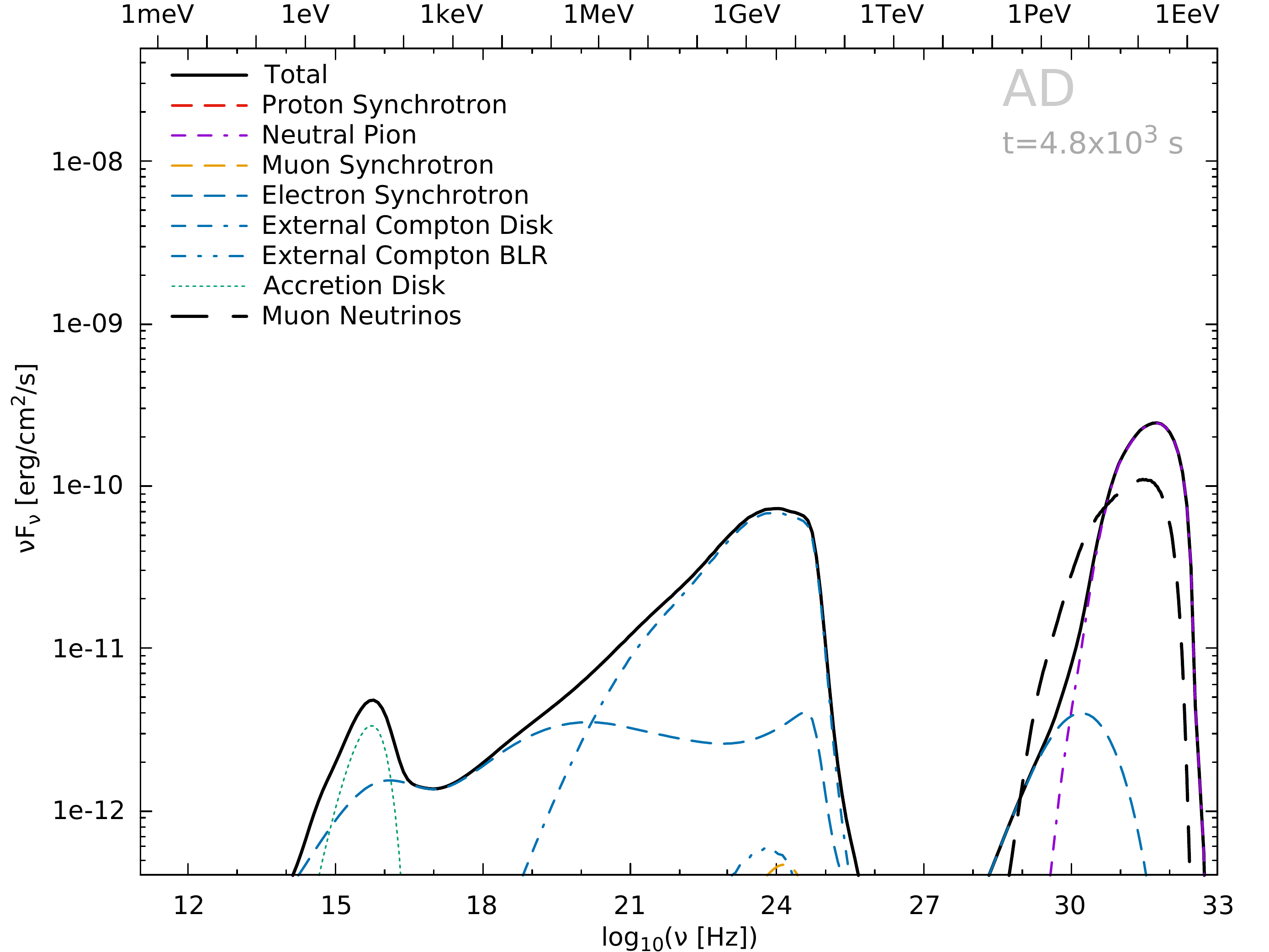}}
\end{minipage}
\hspace{\fill}
\begin{minipage}{0.49\linewidth}
\centering \resizebox{\hsize}{!}
{\includegraphics{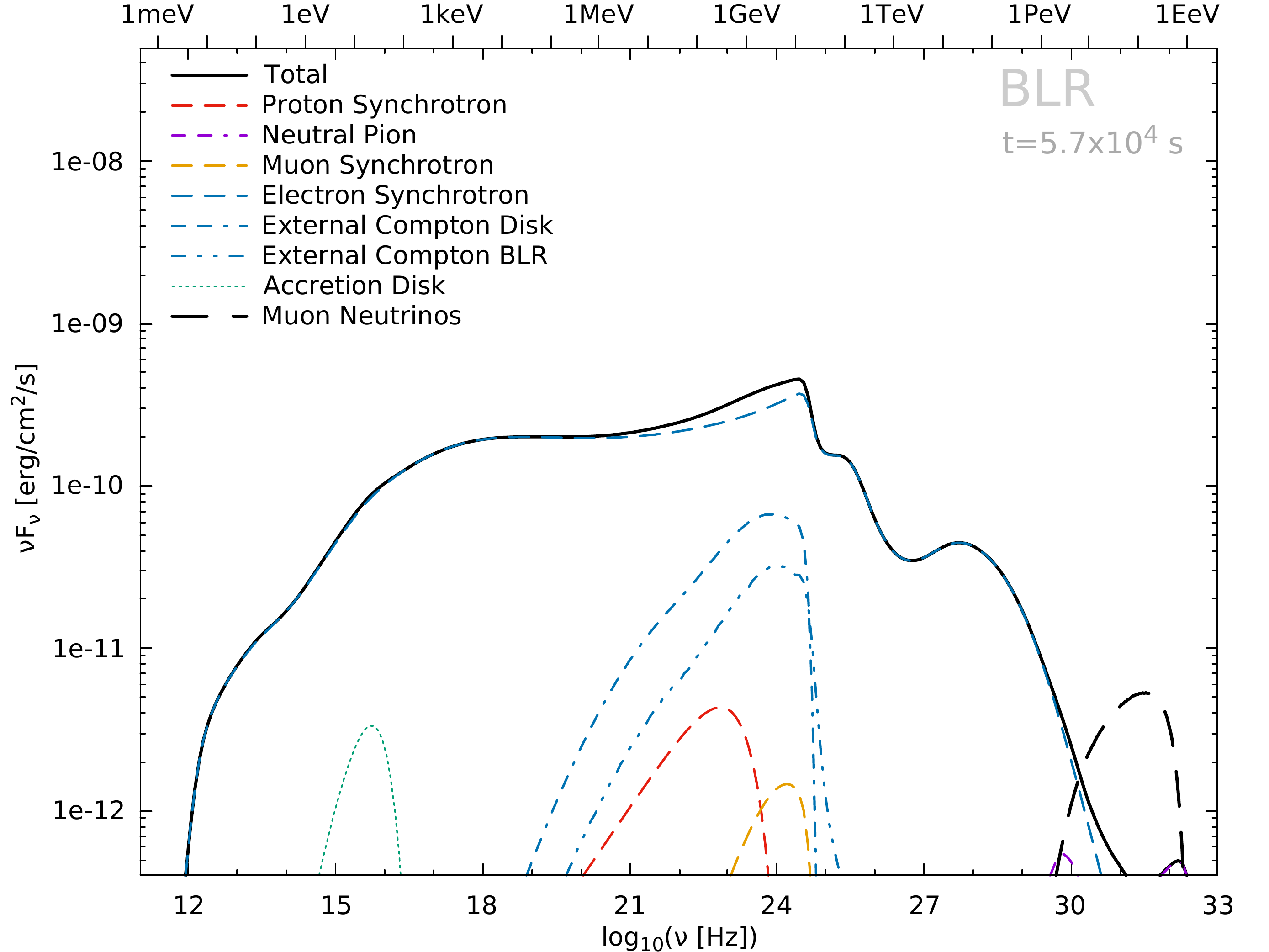}}
\end{minipage}
\newline
\begin{minipage}{0.49\linewidth}
\centering \resizebox{\hsize}{!}
{\includegraphics{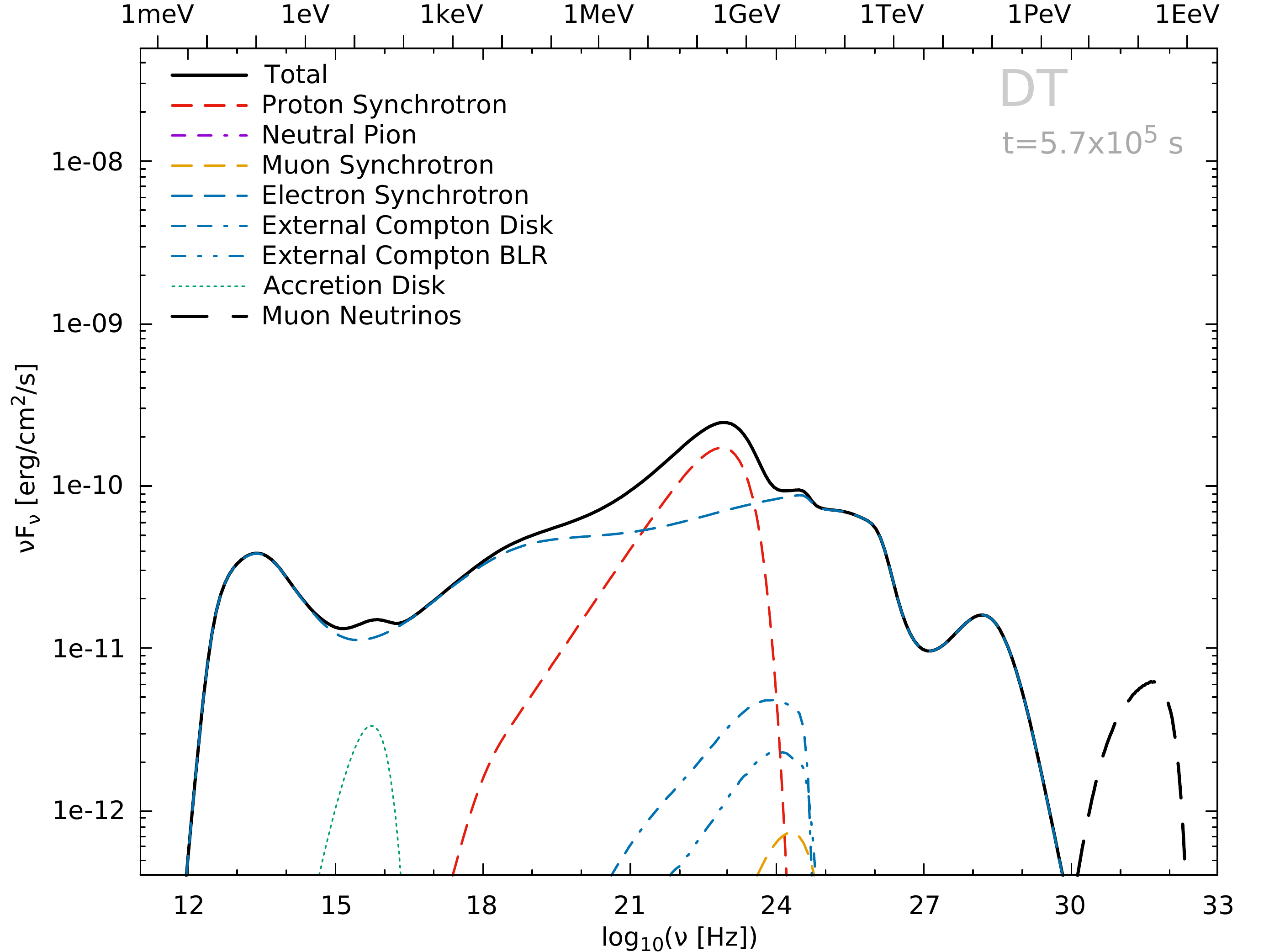}}
\end{minipage}
\hspace{\fill}
\begin{minipage}{0.49\linewidth}
\centering \resizebox{\hsize}{!}
{\includegraphics{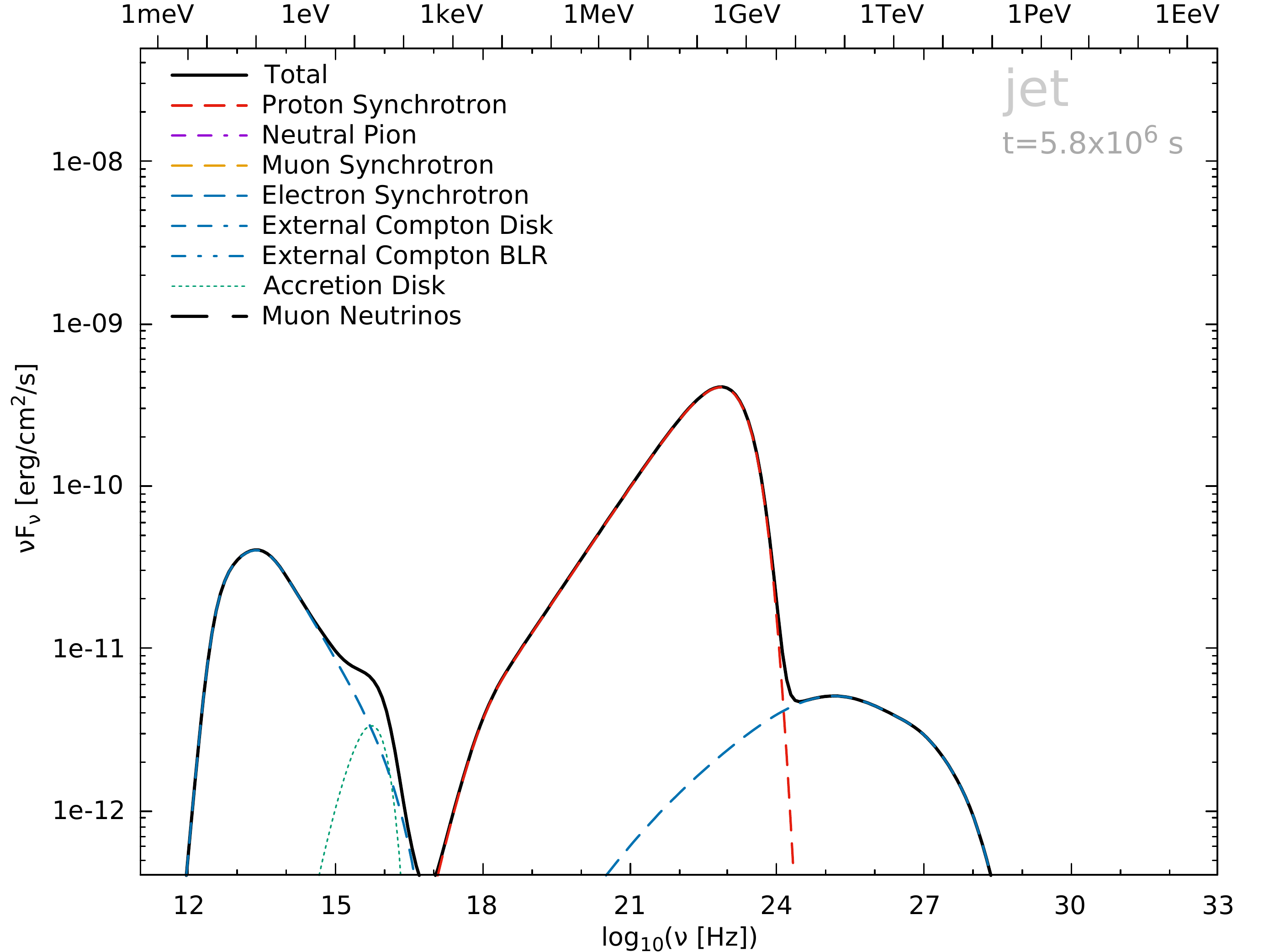}}
\end{minipage}
\caption{Same as Fig.~\ref{fig:spe_ste}, but for a moving blob. 
In each panel, the time in the comoving frame is given that has passed since the launch.  
\label{fig:spe_mot}}
\end{figure}   
\begin{figure}[tbh]
\begin{minipage}{0.49\linewidth}
\centering \resizebox{\hsize}{!}
{\includegraphics{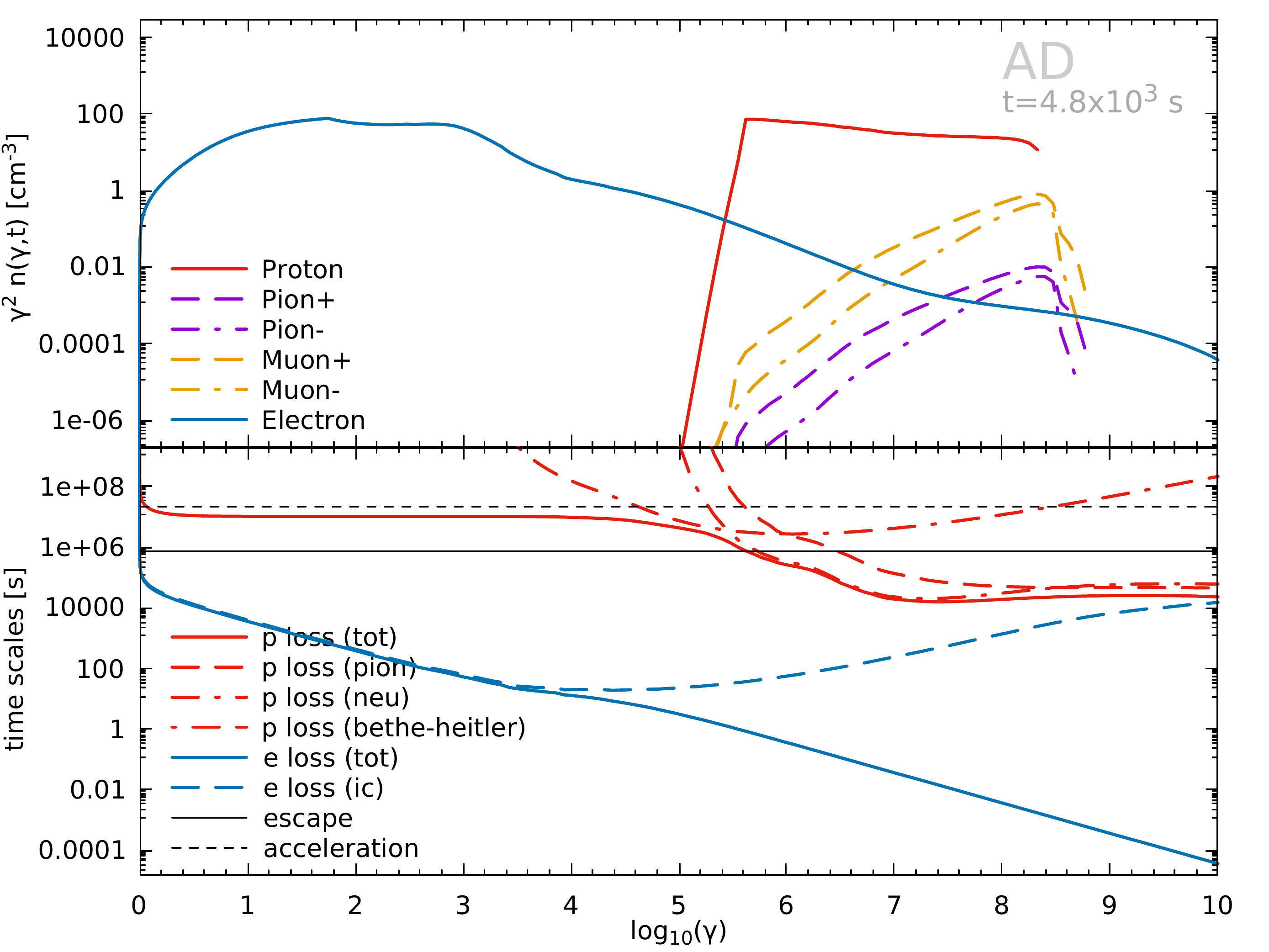}}
\end{minipage}
\hspace{\fill}
\begin{minipage}{0.49\linewidth}
\centering \resizebox{\hsize}{!}
{\includegraphics{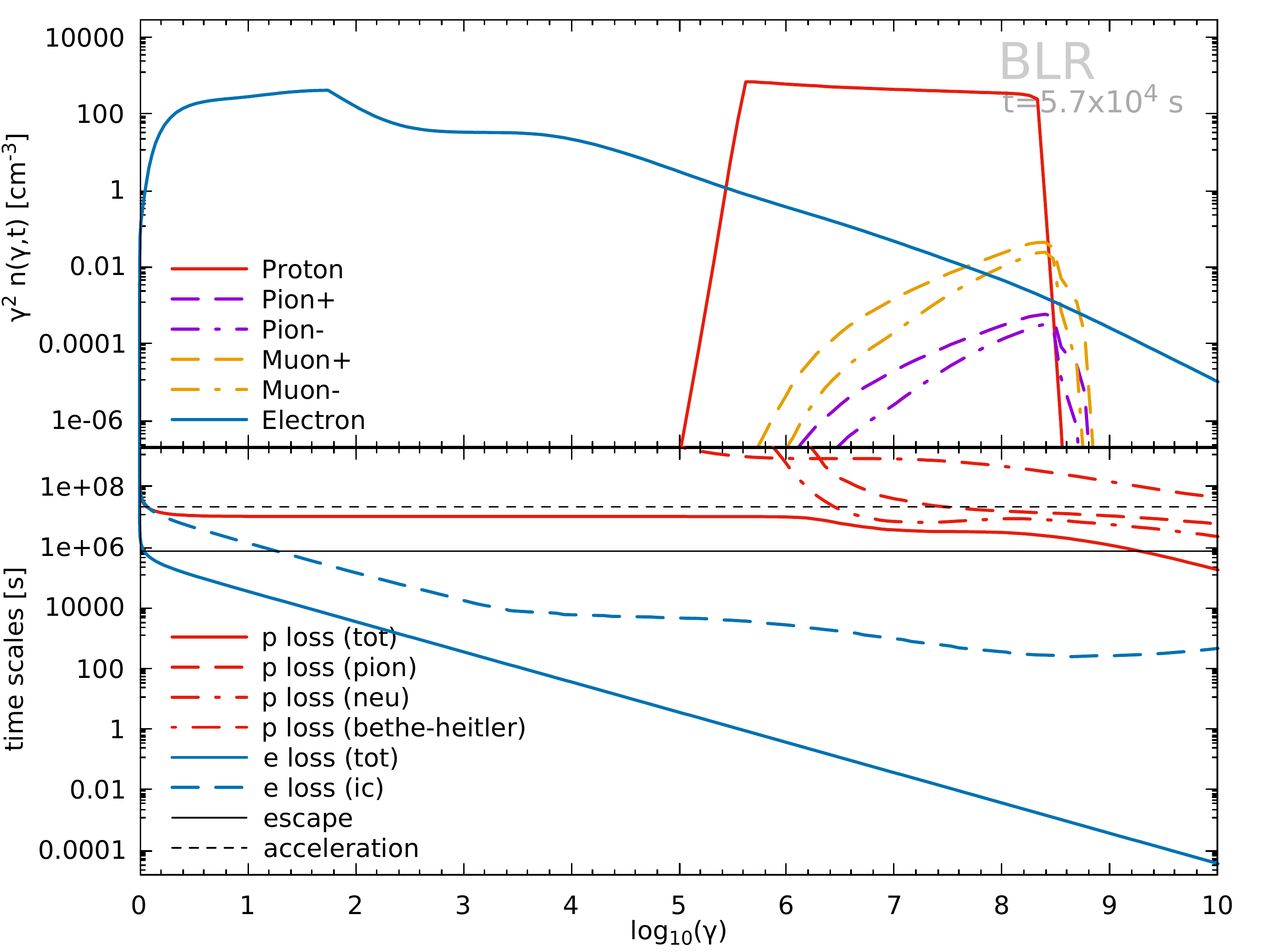}}
\end{minipage}
\newline
\begin{minipage}{0.49\linewidth}
\centering \resizebox{\hsize}{!}
{\includegraphics{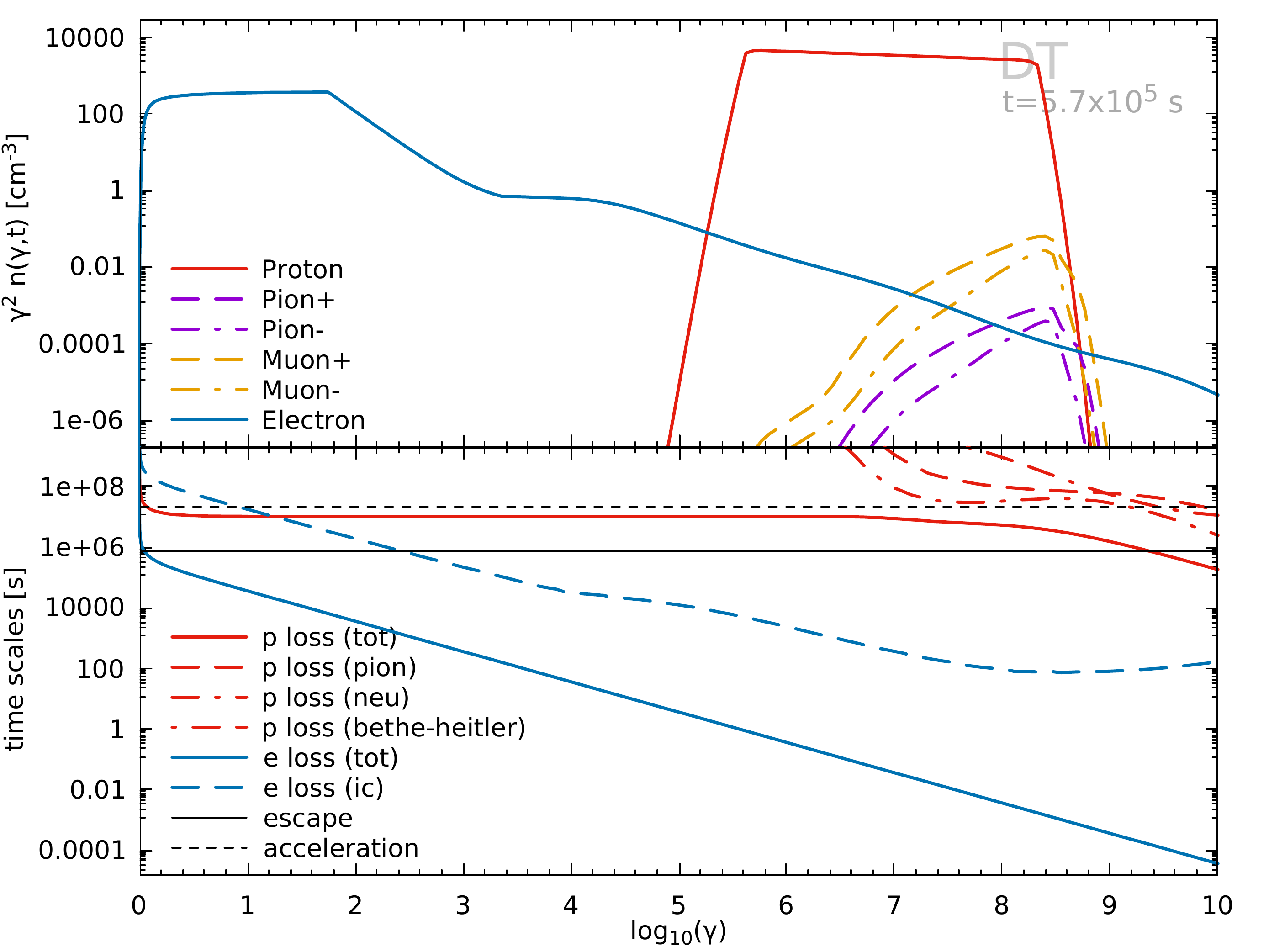}}
\end{minipage}
\hspace{\fill}
\begin{minipage}{0.49\linewidth}
\centering \resizebox{\hsize}{!}
{\includegraphics{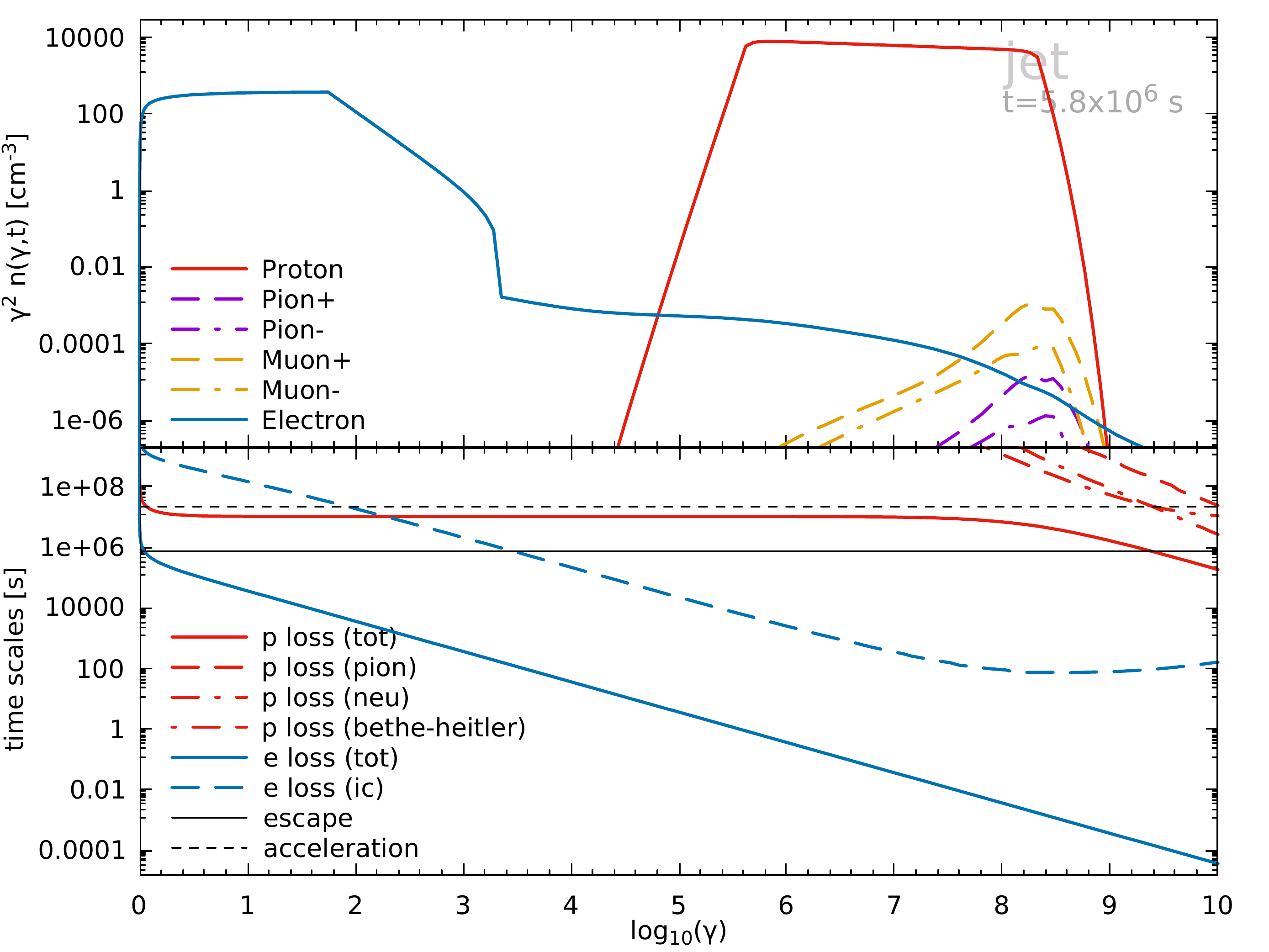}}
\end{minipage}
\caption{Same as Fig.~\ref{fig:par_ste}, but for a moving blob as in Fig.~\ref{fig:spe_mot}. In each panel, the time in the comoving frame is given that has passed since the launch.  
\label{fig:par_mot}}
\end{figure}   
\begin{figure}[tbh]
\begin{minipage}{0.49\linewidth}
\centering \resizebox{\hsize}{!}
{\includegraphics{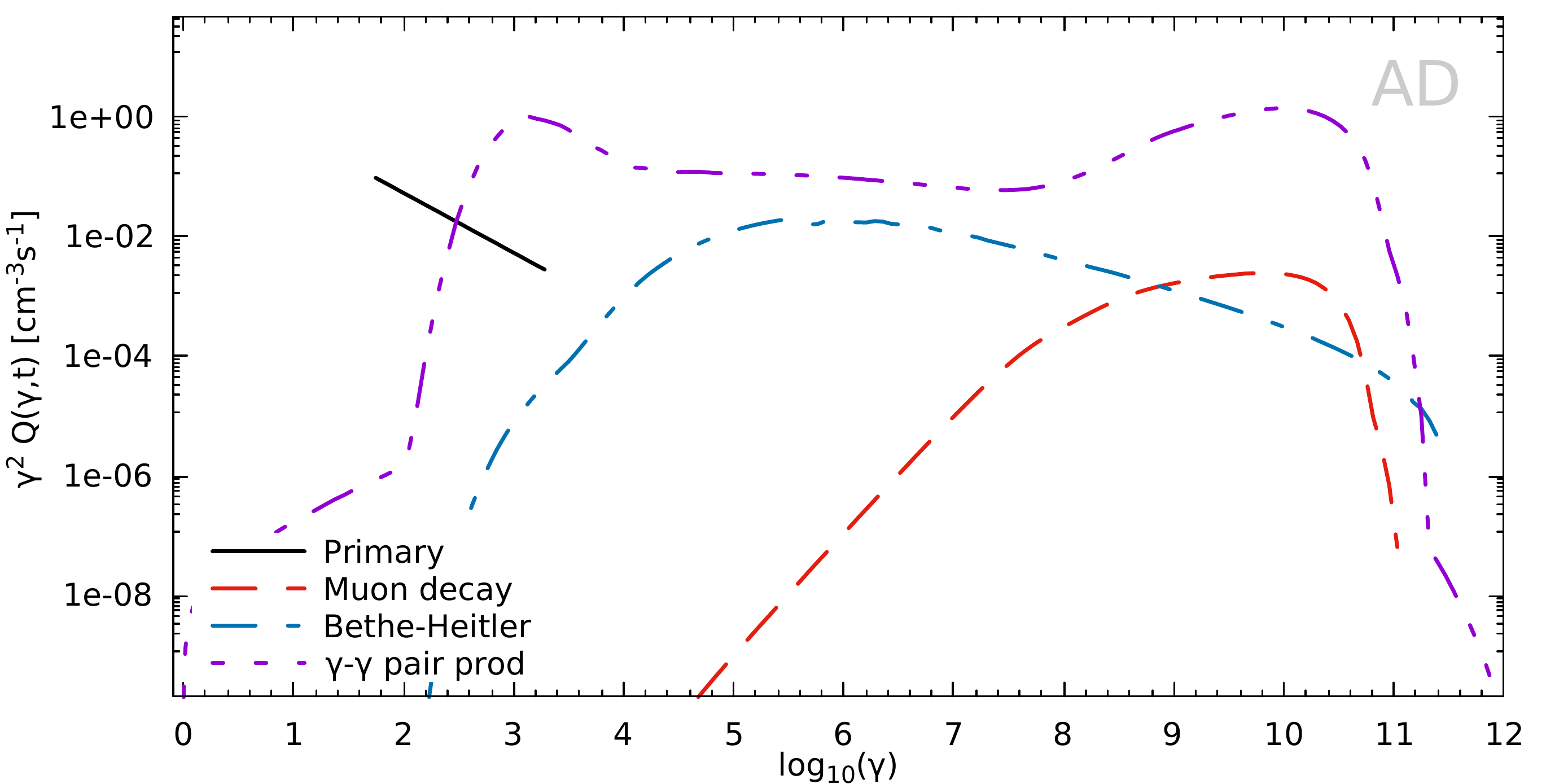}}
\end{minipage}
\hspace{\fill}
\begin{minipage}{0.49\linewidth}
\centering \resizebox{\hsize}{!}
{\includegraphics{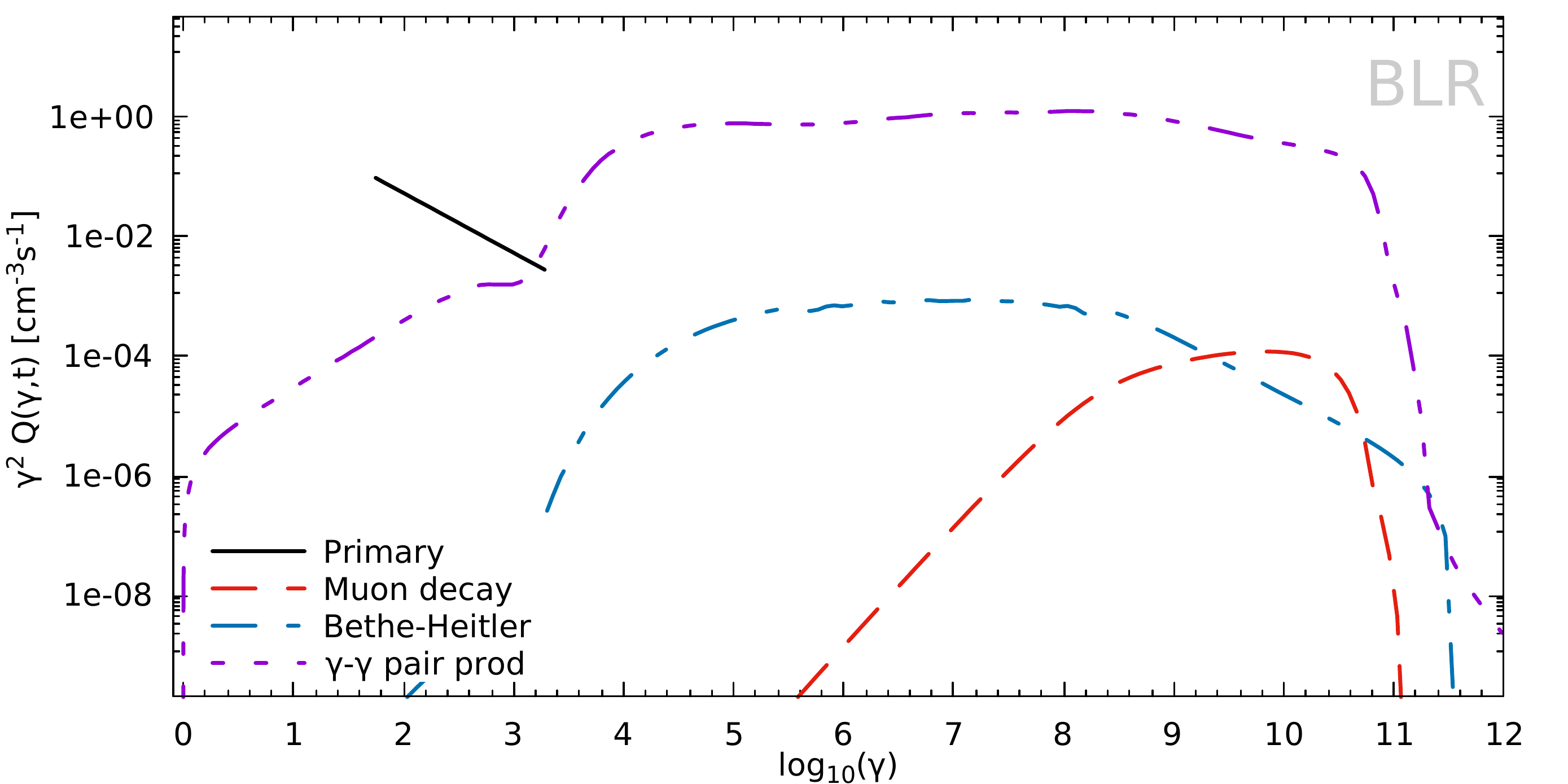}}
\end{minipage}
\newline
\begin{minipage}{0.49\linewidth}
\centering \resizebox{\hsize}{!}
{\includegraphics{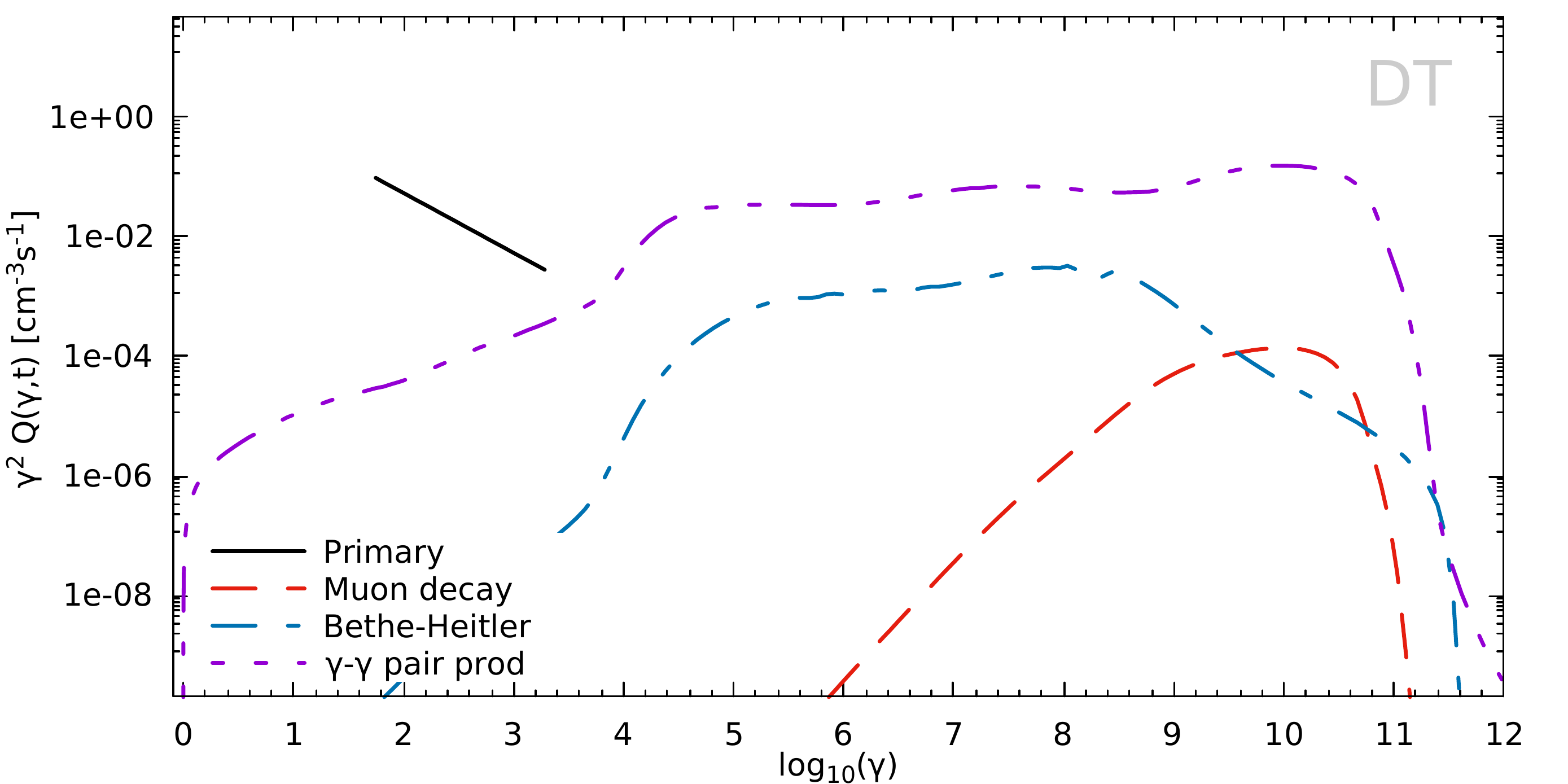}}
\end{minipage}
\hspace{\fill}
\begin{minipage}{0.49\linewidth}
\centering \resizebox{\hsize}{!}
{\includegraphics{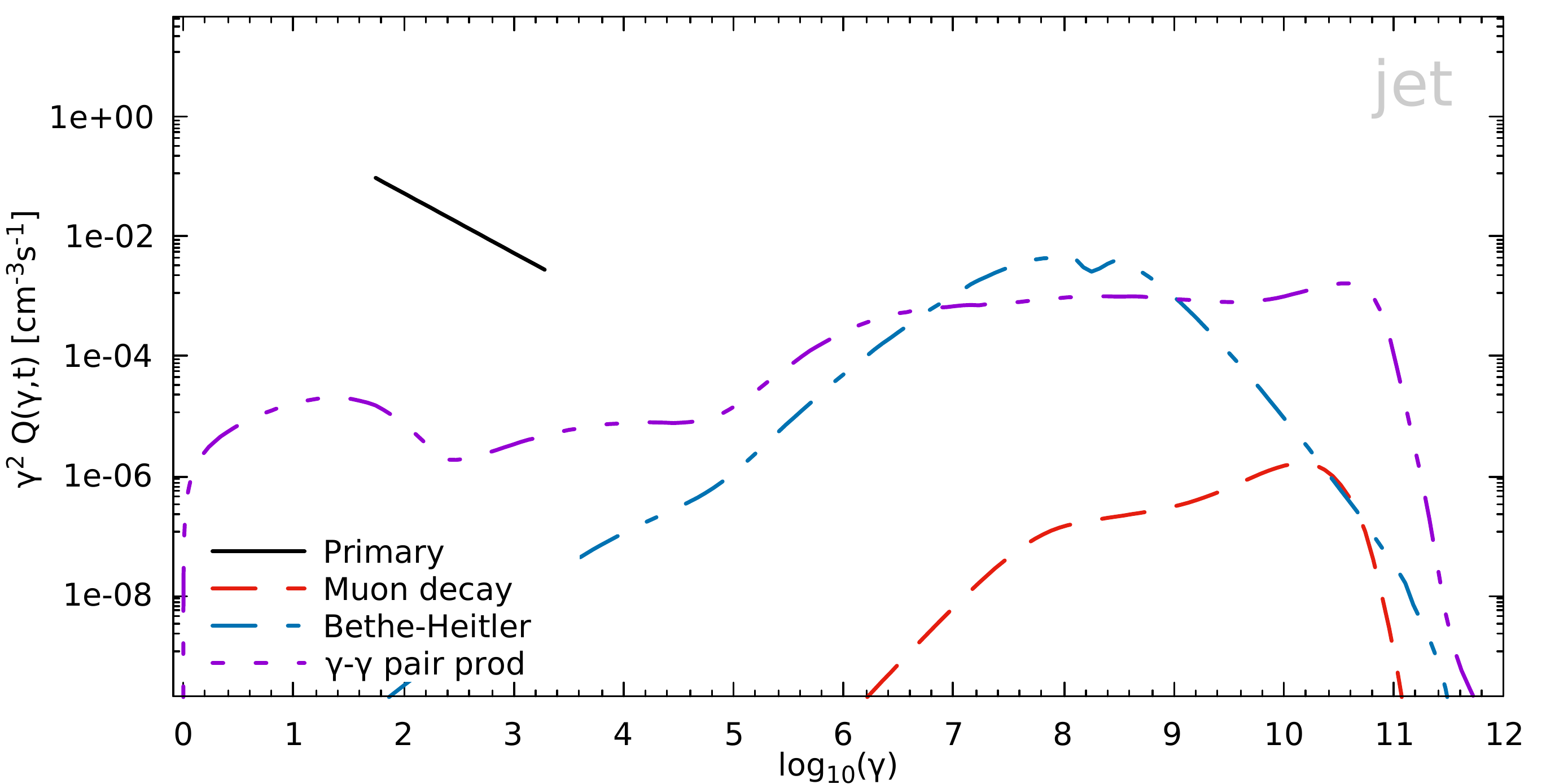}}
\end{minipage}
\caption{Same as Fig.~\ref{fig:inj_ste}, but for a moving blob as in Fig.~\ref{fig:spe_mot}. 
\label{fig:inj_mot}}
\end{figure}   
\begin{specialtable}[tbh] 
\small
\caption{Energy densities in particles $u_{\rm par}$ (in erg/cm$^3$) and the ratio $u_B/u_{\rm par}$ of magnetic to particle energy density. The magnetic energy density in all cases is $u_B = 100\,$erg/cm$^3$. The horizontal line separates the steady-state (top) from the moving (bottom) cases.
\label{tab:endens}}
\begin{tabular}{lcc}
\toprule
\textbf{Position} & \textbf{$u_{\rm par}$}	& \textbf{$u_B/u_{\rm par}$}\\
\midrule
AD	& $13.4$	& $7.5$ \\
BLR	& $55.6$	& $1.8$ \\
DT	& $59.1$	& $1.7$ \\
jet	& $59.7$	& $1.7$ \\
\midrule
AD	& $0.38$	& $263$ \\
BLR	& $4.48$	& $22.3$ \\
DT	& $33.3$	& $3.0$ \\
jet	& $59.7$	& $1.7$ \\
\bottomrule
\end{tabular}
\end{specialtable}
Table~\ref{tab:freeparam} provides an overview of the free parameters that have been described in the previous section. The given parameter values are a toy model, which we use to perform a small parameter study. The parameters are based upon the flat spectrum radio quasar 3C~279 \cite{hess193c279}, however a direct data comparison is beyond the scope of this paper.
Instead, we wish to analyze the influence of the external fields on the SED and the particle distributions. We chose four locations: close to the AD ($z_0=1\E{16}\,$cm), within the BLR ($z_0=1\E{17}\,$cm), within the DT  ($z_0=1\E{18}\,$cm), and outside the external fields (referred to as ``jet'', $z_0=1\E{19}\,$cm). All other parameters remain unchanged including the radius and the magnetic field of the emission region. This highlights that these are indeed toy models meant to study the influence of the external fields without any degeneracies introduced by varying other parameters.


The result is shown in Figs.~\ref{fig:spe_ste}, \ref{fig:par_ste} and \ref{fig:inj_ste}. While the SEDs are transformed into the observer's frame, the photon spectra are shown as they leave the emission region in the jet. The internal \g-\g\ absorption processes are fully considered (the corresponding optical depth $\tau_{\gamma\gamma}$ is shown in Fig.~\ref{fig:abs} left), however we do not show the additional absorption of \g\ rays while traveling through the photon field of the host galaxy (namely BLR and DT) or through the cosmological photon fields (extragalactic background light and CMB). Any of these photon fields could additionally (and severely) attenuate the photon flux above 10\,GeV. These absorption processes are, however, not important for the conclusions of this study.

Close to the AD, the external fields are very intense, and are further enhanced through the large chosen bulk Lorentz factor of $50$. In turn, the cooling of protons through proton-photon interactions is very strong (Fig.~\ref{fig:par_ste}), as indicated by the cooling time scales being dominated by pion production (indicated by the ``pion'' and ``neutron'' loss channels) at Lorentz factors $\g>10^5$. This severely influences the proton distribution function and results in negligible proton synchrotron emission. The strong pion production, which can also be seen in the SED (Fig.~\ref{fig:spe_ste}) through the neutral pion bump at PeV energies, results in a significant production of muons and highly relativistic electrons (Fig.~\ref{fig:inj_ste}) with Lorentz factors $\g>10^{10}$. Similarly, highly energetic electrons are also injected through Bethe-Heitler pair production. These electrons produce \g\ rays through synchrotron emission, as well as through IC emission for lower-energetic electrons. The \g\ rays are absorbed through \g-\g\ pair production with all photon fields that permeate the emission region. The strength of the \g-\g\ absorption is shown in the left panel of Fig.~\ref{fig:abs}, and manifests itself in Fig.~\ref{fig:spe_ste} by the significant flux suppression at energies above 10\,GeV. In turn, a strong electron-positron cascade is initiated. This results in an electron distribution which is dominated by secondaries (Fig.~\ref{fig:inj_ste}). The resulting electron synchrotron flux (Fig.~\ref{fig:spe_ste}) extends through almost the entire frequency range destroying the familiar double-hump shape in the SED. The peak of the flux at \g\ rays stems from IC scattering of AD photons. 

Within the BLR, the proton cooling is drastically reduced at high Lorentz factors with cooling time scales being longer than the escape time scale of particles at all (relevant) energies (Fig.~\ref{fig:par_ste}). Unlike in the AD case, where the proton distribution cuts off sharply at $\gamma_{\rm max,p}$, in this case (and the following cases) the proton distribution extends beyond the injection cut-off because of the (re-)acceleration terms present in Eq.~(\ref{eq:fpgen}). The change in the spectral shape between the AD and BLR cases allows for an enhanced proton synchrotron emission in the BLR case, influencing the SED at GeV energies (Fig.~\ref{fig:spe_ste}). While pion and Bethe-Heitler pair production are reduced compared to the AD case, the pair cascade is still very significant (Fig.~\ref{fig:inj_ste}) because of \g-\g\ pair production (Fig.~\ref{fig:abs} left). While the process is less severe than in the AD case, the secondaries still dominate the electron distribution (Fig.~\ref{fig:par_ste}), and produce synchrotron emission beyond PeV energies. In the BLR case IC emission is negligible.

This trend continues in the DT case, as the cascade weakens (Fig.~\ref{fig:inj_ste} and Fig.~\ref{fig:abs} left) and the more familiar double-humped SED emerges (Fig.~\ref{fig:spe_ste}). At UV energies in the SED, a minor contribution from the AD itself is visible. The \g-ray peak is  dominated by proton synchrotron emission, even though the secondary electron synchrotron emission still dominates at X-ray and TeV energies. The neutral pion bump is below the shown flux scale indicating the reduced interaction of protons with photons. In fact, the protons are completely in a slow-cooling regime (Fig.~\ref{fig:par_ste}).

Lastly, the emission region is located outside the external photon fields in the ``jet'' case. While secondary pairs are still being produced (Fig.~\ref{fig:inj_ste}), their number is low (Fig.~\ref{fig:par_ste}) due to low absorption (Fig.~\ref{fig:abs} left), and their flux contribution only shows around photon energies on the TeV-scale, but at relatively low flux values (Fig.~\ref{fig:spe_ste}). Apart from that, the SED is dominated by synchrotron emission of protons at X- and \g\ rays, and primary electrons in the optical domain. Both peaks are cleanly separated. The AD itself is clearly visible in the UV range as a big blue bump. 

The changes in the cooling strength can also be seen in the energy densities of the particles, which are given in Tab.~\ref{tab:endens}. The particle energy densities are always dominated by protons (by several orders of magnitude compared to the electrons in most cases). The strong cooling in the AD case results in a low particle energy density, while the reduced cooling in the other cases results in increased and comparable energy densities. Given the constant value of the magnetic field in all cases, the ratio of magnetic to particle energy density decreases from case to case but is always larger than unity.

The different cases are also manifested in the emerging neutrino spectra. With the weakening production of pions and muons from case to case, the flux of neutrinos also decreases and drops below the scale of the plots in the ``jet'' case. The AD case produces not just the highest neutrino flux, but also a different neutrino spectral shape than the other cases with a flat maximum (or mildly double-humped structure) over almost 3 orders of magnitude in energy. In the BLR and DT case, the neutrino spectra show a single peak at about 100\,PeV. Interestingly, all three cases would be detectable with the future IceCube-Gen2 instrument \cite{icecubegentwo21}. However, the unrealistic SEDs -- especially in the AD and BLR cases -- make it seem unlikely that neutrinos could be observed from a blazar -- at least, under this simple set-up.


For the examples discussed above, we have used a bulk Lorentz factor of $50$. Hence, if the emission region were moving, it would cover a lot of space in a relatively short amount of time because of Lorentz contraction: $\hat{z}=z_0+\Gamma \beta_{\Gamma}c t$, where $t$ is the time since launch in the comoving frame, and $\hat{z}$ is the location of the emission region in the host galaxy frame. In turn, the external fields, and thus the conditions within the emission region may change quickly. We try to analyze this, by letting the emission region flow from the base (placed at six times the Schwarzschild radius (innermost stable circular orbit) of the black hole) downstream through the jet. 

As before, none of the other parameters change implying that also the primary injection of protons and electrons continues with the same rate $Q$ and spectral shape throughout the simulation. This assumes a quasi-instantaneous acceleration of particles \cite{baringboettcher19}, as well as a continuous supply. This is not realistic, as the acceleration of particles also takes time \cite{weidingerspanier15}.
Additionally, neither the magnetic field $B$ nor the radius $R$ vary. While the radius of the emission region may not expand as rapidly as the larger jet structure that surrounds it, it expands nonetheless while it travels through the jet \cite{boulamastichiadis21} given the high energy densities in the emission region. While recent observational results \cite{hess193c279,hess+21pks1510} indicate compact emission regions beyond the BLR, and maybe even at tens of parsecs from the black hole, it is not clear whether these are indeed moving emission regions originating close to the black hole or turbulent cells within a larger flaring region.
Similarly, while a high magnetic field can be expected close to the black hole, the expansion of the emission region causes a drop of the magnetic field with increasing distance. 
%
These considerations highlight once more the toy character of this study. 
Applying such parameter changes are interesting avenues for future studies beyond the scope of this paper. 

Having obtained the full journey of the emission region through the jet, we extract the SEDs and particle distributions at the same distances as in the steady-state cases \footnote{Given the finite time resolution in the simulation, we extract the SEDs and particle distribution at the time step closest to the respective distances of the steady-state models (AD: $9.83\E{15}\,$cm, BLR: $9.75\E{16}\,$cm, DT: $9.64\E{17}\,$cm, jet: $9.75\E{18}\,$cm). In order to save computation time, while also properly resolving the initial steps within the BLR, we use an adaptive time step of $\Delta t_i = 1\E{3+i/20}$, where $i$ is the number of the step. This ensures reasonable accuracy and resolution, and also explains why the time values given in Figs.~\ref{fig:spe_mot} and \ref{fig:par_mot} are not simple increases by a factor 10, as one would expect. We believe that this is a reasonable trade-off.}. 
The results are shown in Figs.~\ref{fig:spe_mot}, \ref{fig:par_mot}, and \ref{fig:inj_mot}, while the optical depth due to \g-\g pair production is shown in the right panel of Fig.~\ref{fig:abs}. We note that any times and time scales discussed below are in the comoving frame. 

The changes to the SEDs and the particle spectra are profound. The emission region has passed the AD position after merely $5\,$ks. The bright external photon fields cause proton-photon interactions producing a significant amount of pions (Fig.~\ref{fig:par_mot}), which decay into photons or muons and pairs. In fact, Fig.~\ref{fig:inj_mot} shows that the injection of pairs from muon decay is almost at the level as in the steady state (Fig.~\ref{fig:inj_ste}), but Bethe-Heitler produced pairs are about 2 orders of magnitude below. Similarly, \g-\g\ pair production is below the steady-state level, because the internal photon fields (Fig.~\ref{fig:spe_mot}) have not yet been fully developed. In turn, the optical depth due to \g-\g\ pair production (Fig.~\ref{fig:abs} right) is not at the steady-state level -- merely the absorption caused by external fields is fully present. One consequence is the reduced absorption at PeV photon energies allowing for a very strong flux in the neutral pion bump (Fig.~\ref{fig:spe_mot}). The ``under-development'' of the internal photon fields is a consequence of the low electron and proton densities (Fig.~\ref{fig:par_mot}) compared to the steady-state values. The consequence is the absence of the ``nominal'' electron synchrotron bump in the infrared domain. The \g\ rays are dominated by IC scattering of AD photons -- though orders of magnitude below the steady-state case.

The situation only changes mildly until the BLR position is reached after $5.7\E{4}\,$s. This is still less than the escape times of photons ($2\E{5}\,$s) and particles ($7.5\E{5}\,$s). Therefore, particle and photon densities continue to increase. The spectral shape of the SED shown in Fig.~\ref{fig:spe_mot} is somewhat similar to the steady-state case (Fig.~\ref{fig:spe_ste}), but at a factor of a few reduced in flux. There are a few more details where SEDs differ. In the \g-ray domain, Fig.~\ref{fig:spe_mot} shows contributions from IC scattering of both the AD and the BLR. Comparing the IC/AD spectra of the top panels in Fig.~\ref{fig:spe_mot}, one notices the similarity between them. Given that not even one light-crossing time scale has passed since the launch, the photons produced below have not yet vanished from the emission region, and therefore continue to contribute to the SED even though the IC/AD production has much reduced at this distance. The IC/BLR spectrum shows a different spectral shape and a higher flux than in the steady-state case, which can be attributed to the slightly different shapes in the electron distributions (Fig.~\ref{fig:par_mot} vs. Fig.~\ref{fig:par_ste}). These are a consequence of the reduced \g-\g\ pair production at lower energies (Fig.~\ref{fig:inj_mot}). As the protons have also not reached the steady-state density, their synchrotron flux is reduced compared to the steady state, while pion and muon production are similarly reduced. Most notably, the neutral pion decay flux is barely visible at PeV to EeV energies in Fig.~\ref{fig:spe_mot} -- a reduction of about an order of magnitude compared to the steady state.


After $5.7\E{5}\,$s, the emission region has reached the DT position. The time the emission region has traveled, is now comparable to the escape time scales of light and particles. In turn, the SED in Fig.~\ref{fig:spe_mot} is almost equal to the steady-state case (Fig.~\ref{fig:spe_ste}), except for a reduced peak \g-ray flux by a factor of a few. This can be attributed to the still lower number of protons compared to the steady state resulting in an equally reduced proton synchrotron flux. This is also coupled to the low efficiency of proton-synchrotron emission implying that the flux needs more time to build compared to the electron synchrotron flux, which is basically instantaneous -- cf. the cooling time scales in Fig.~\ref{fig:par_mot}, where electron synchrotron cooling is faster than basically any other time scale (including the travel time), while the proton synchrotron cooling only dominates at energies beyond the cut-off of the proton distribution. The IC/AD and IC/BLR components visible in the SED (Fig.~\ref{fig:spe_mot}) exhibit a flux about an order of magnitude below the flux at the BLR position. This corresponds very well to an exponential decay, as the photons leave the emission region without being replenished. 


The following, relatively long cruise towards the ``jet'' position (reached after $5.8\E{6}\,$s) allows for the near-complete relaxation of the emission region towards the steady state that was obtained above. At this position, SED and particle distributions are practically equal to the steady-state case.

The particle energy densities change considerably from position to position because of the accumulation of relativistic particles in the emission region. This is the reason why the particle energy density at the AD position is about a factor $35$ lower than in the steady state case. This accumulation of particles continues through the other position increasing the particle energy density along the way until the jet position, where the previous steady-state value is obtained. Similarly, the ratio of energy densities is initially very large and decreases on the way out.

The neutrino spectra shown in Fig.~\ref{fig:spe_mot} indicate as well that the interactions and distributions require time to unfold. While at the AD position lots of neutrinos are produced, their flux is a factor of a few below the steady-state flux. At the BLR position the flux reduction is almost an order of magnitude (similar to the pion flux), while it is closer to the steady-state flux at the DT position. At the ``jet'' position, the neutrino flux is much reduced as in the steady-state case.


\section{Discussion and conclusion}
The results of the toy study presented in this paper clearly show the importance of the external fields in case of the presence of relativistic protons in the jet. Their influence on the particle evolution is significant resulting in very different steady-state SEDs at different positions in the jet. Especially at locations within the BLR, the familiar double-humped SED structure is destroyed. At the DT position, the spectrum is already comparable to ``standard'' blazar SEDs, while the ``jet'' position outside the external fields provides the cleanest separation between the low-energy and the high-energy bump.


The situation changes entirely when the motion of the emission region is taken into account. The relatively long source time scales (particle and photon accumulation, interactions, escape) compared to the fast speed imply that the external conditions change too fast for the emission region to adapt even until the edge of the DT. Only on ``jet'' scales, the previous steady state is fully recovered. This, of course, is a consequence of the choice of $\Gamma=50$, which is a rather extreme value. Lower values on the order of $\Gamma\sim 10$ could change the situation -- especially as it would also significantly reduce the energy density of the external fields within the emission region. Steady-state solutions might be achieved at positions much closer to the black hole. Testing this, and the other potential changes to the model parameters as described above, is however beyond the scope of this paper.

Within the model parameters used in this toy study, the production of neutrinos depends strongly on the external fields with practically none produced at the ``jet'' position. While different parameter sets of the emission region might produce better SED shapes at positions within the external photon fields, it corroborates the results obtained by other authors \cite{gao+19,cerruti+19,rbb19} that it is difficult to reconcile the neutrino and photon observations within a one-zone model.

To conclude, the production of neutrinos in a blazar jet in reasonable quantities remains a challenge, as the requirement for a reasonably dense soft photon field -- in order to produce the required pions -- also supports the pair cascade through \g-\g\ absorption and Bethe-Heitler pair production. The intrusion of a gas cloud or a star into the jet \cite{bpb12,zacharias+17} might provide sufficient numbers of cold protons for direct proton-proton interactions \cite{hoerbe+20}, but the consequences (efficiency of the process, developing pair cascade, etc) would also need further studies.


\funding{The author acknowledges postdoctoral financial support from LUTH, Observatoire de Paris.}

\dataavailability{The OneHaLe code is still under development and therefore not yet meant for public use. However, the code can be shared upon reasonable request to the author.} 

\acknowledgments{I am eternally grateful to Prof. Dr. Reinhard Schlickeiser for his supervision and guidance from the Bachelor thesis up to the Ph.D. thesis and beyond. Thank you very much for all the guidance along the way -- and for all the travel destination that you have made possible. Because of you, I realized quickly that this is indeed what I want to do -- being a researcher, I mean. \\
I am also grateful for stimulating discussions with Anita Reimer, Andreas Zech, Catherine Boisson, Markus B\"ottcher, and Chris Diltz, which helped in creating and improving the code. I would like to thank the referees for constructive reports that helped a lot to improve the paper. 
Simulations for this paper have been performed on the TAU-cluster of the Centre for Space Research at North-West University, Potchesftroom, South Africa.}

\conflictsofinterest{The author declares no conflict of interest.}

\end{paracol}
\reftitle{References}



\externalbibliography{yes}
\bibliography{onehale}

\end{document}